\documentclass[a4paper,fleqn,usenatbib]{mnras}
\usepackage[T1]{fontenc}
\usepackage{ae,aecompl}

\usepackage{graphicx}	
\usepackage{amsmath}	
\usepackage{amssymb}	
\usepackage[utf8]{inputenc}
\usepackage{epsfig}
\usepackage{url}
\usepackage{capt-of}
\usepackage{multicol}
\usepackage{epstopdf}
\usepackage{caption}
\usepackage{textcomp}
\usepackage[usenames,dvipsnames]{color}
\usepackage{float,lscape}
\usepackage{pdfpages}
\usepackage{pdflscape}
\usepackage{afterpage}
\usepackage{rotating}
\usepackage{arydshln}

\title[The Rotation Curves of Star-forming Galaxies]{The Shapes of the Rotation Curves of Star-forming Galaxies Over the Last $\mathbf{\approx}$$\mathbf{10}$ Gyr}

\author[Tiley et al.]{Alfred L.\ Tiley,$^{1\dagger}$ A. M. Swinbank,$^{1}$ C. M. Harrison,$^{2}$ Ian Smail,$^{1}$ O. J. Turner,$^{3}$ 
\newauthor  M. Schaller,$^{4,5}$ J. P. Stott,$^{6}$ D. Sobral,$^{6}$ T. Theuns,$^{4}$ R. M. Sharples,$^{7,1}$ S. Gillman,$^{1}$ 
\newauthor R. G. Bower,$^{4,1}$ A. J. Bunker,$^{8,9}$ P. Best,$^{3}$ J. Richard,$^{10}$ Roland Bacon,$^{10}$ M. Bureau,$^{8,11}$   
\newauthor M. Cirasuolo,$^{2}$ G. Magdis$^{12,13}$
\\
$^{1}$Centre for Extragalactic Astronomy, Department of Physics, Durham University, South Road, Durham, DH1 3LE, U.K.\\
$^{2}$European Southern Observatory, Karl-Schwarzchild-Str. 2, 85748 Garching b. M{\"u}nchen, Germany\\
$^{3}$ Institute for Astronomy, University of Edinburgh, Royal Observatory, Edinburgh EH9 3HJ\\
$^{4}$Institute for Computational Cosmology, Durham University, South Road, Durham, DH1 3LE, U.K.\\
$^{5}$ Leiden Observatory, Leiden University, PO Box 9513, 2300 RA, Leiden, the Netherlands\\
$^{6}$Department of Physics, Lancaster University, Lancaster, LA1 4YB, U.K.\\
$^{7}$Centre for Advanced Instrumentation, Department of Physics, Durham University, South Road, Durham, DH1 3LE, U.K.\\
$^{8}$Sub-dept. of Astrophysics, Department of Physics, University of Oxford, Denys Wilkinson Building, Keble Road, Oxford, OX1 3RH, U.K.\\
$^{9}$Affiliate Member, Kavli Institute for the Physics and Mathematics of the Universe, 5-1-5 Kashiwanoha, Kashiwa, 277-8583, Japan\\
$^{10}$Univ Lyon, Univ Lyon1, Ens de Lyon, CNRS, Centre de Recherche Astrophysique de Lyon UMR5574, F-69230, Saint-Genis-Laval, France\\
$^{11}$Yonsei Frontier Lab and Department of Astronomy, Yonsei University, 50 Yonsei-ro, Seodaemun-gu, Seoul 03722, Republic of Korea\\
$^{12}$Cosmic DAWN Centre, Niels Bohr Institute, University of Copenhagen, Juliane Mariesvej 30, 2100, Copenhagen, Denmark \\
$^{13}$Institute for Astronomy, Astrophysics, Space Applications and Remote Sensing, National Observatory of Athens, GR-15236 Athens, Greece\\
$^{\dagger}$E-mail: alfred.l.tiley@durham.ac.uk
}

\date{Accepted XXX. Received YYY; in original form ZZZ}

\pubyear{2018}

\begin{document}
\label{firstpage}
\pagerange{\pageref{firstpage}--\pageref{lastpage}}
\maketitle

\begin{abstract}
We analyse maps of the spatially-resolved nebular emission of $\approx$1500 star-forming galaxies at $z\approx0.6$--$2.2$ from deep KMOS and MUSE observations to measure the average shape of their rotation curves. We use these to test claims for declining rotation curves at large radii in galaxies at $z\approx1$--$2$ that have been interpreted as evidence for an absence of dark matter. We show that the shape of the average rotation curves, and the extent to which they decline beyond their peak velocities, depends upon the normalisation prescription used to construct the average curve. Normalising in size by the galaxy stellar disk-scale length after accounting for seeing effects ($R_{\rm{d}}^{\prime}$), we construct stacked position-velocity diagrams that trace the average galaxy rotation curve out to $6R_{\rm{d}}^{\prime}$ ($\approx$13\, kpc, on average). Combining these curves with average H{\sc i} rotation curves for local systems, we investigate how the shapes of galaxy rotation curves evolve over $\approx$10 Gyr. The average rotation curve for galaxies binned in stellar mass, stellar surface mass density and/or redshift is approximately flat, or continues to rise, out to at least $6R_{\rm{d}}^{\prime}$. We find a trend between the outer slopes of galaxies' rotation curves and their stellar mass surface densities, with the higher surface density systems exhibiting flatter rotation curves. Drawing comparisons with hydrodynamical simulations, we show that the average shapes of the rotation curves for our sample of massive, star-forming galaxies at $z\approx0$--$2.2$ are consistent with those expected from $\Lambda$CDM theory and imply dark matter fractions within $6R_{\rm{d}}$ of at least $\approx60$ percent.
\end{abstract}




\section{Introduction}
\label{sec:intro}

Galaxy rotation curves, that describe galaxies' circular velocity as a function of galactocentric radius, are very well studied in the local Universe and provide some of the most compelling evidence for the existence of dark matter. When the first systematic measurements of rotation curves were made, the expectation was that they would reveal Keplerian dynamics, with their rotation curves initially rising and then declining at radii beyond that enclosing the visible mass. However, \citet{Hulst:1957} used observations of H{\sc i} 21 cm emission from M31 to show that its rotation curve instead remained approximately flat out to $\approx 25$ kpc, well beyond the radii traced by the stars. \citet{Schmidt:1957} demonstrated that this could be explained if M31 has significant amounts of ``dark'' mass that extends far beyond the spatial extent of the visible matter. The existence of dark matter was later comprehensively recognised in the works of Rubin and others in the 1970s and 1980s \citep[e.g.][]{Rubin:1970,Rubin:1978,Bosma:1978,Rubin:1980,Rubin:1982,Rubin:1985} who demonstrated the ubiquity of flat rotation curves in local spiral galaxies, which can be explained by significant amounts of dark matter residing in a halo which extends well beyond the stars. 

Many subsequent studies have confirmed the results of Rubin et al. \citep[e.g.][]{Catinella:2006,Carignan:2006,deBlok:2008}, each contributing to a now overwhelming body of evidence that shows the ubiquity of dark matter in the cosmos. This evidence includes observations of strong and weak gravitational lensing \citep{Walsh:1979a,Lynds:1986,Tyson:1990}, as well as the discrepancy between the visible mass of clusters galaxies and that deduced via their virial motions \citep{Zwicky:1933} or from the luminosity of X-ray emitting cluster gas in hydrostatic equilibrium \citep[e.g.][]{Fabricant:1980}. Each of which imply the presence of large amounts of unseen matter within galaxies or clusters of galaxies.

Today, the $\Lambda$-cold dark matter paradigm ($\Lambda$CDM), of which dark matter is the corner stone, is a widely accepted description of the framework upon which structure formation is based; it is now well established that dark matter constitutes $\approx$24 percent of the total energy budget of the Universe \citep[e.g.][]{Freedman:2003}, a considerably larger fraction than that of baryonic matter ($\approx4$ percent). Collisionless, cold dark matter forms the framework on which cosmological simulations are based. These simulations witness the initial formation of dark matter halos as small primordial perturbations (in an otherwise smooth matter distribution) are amplified under gravity. Within these halos the baryonic matter collapses to form stars and, later, galaxies. These simulations thus {\it require} dark matter as the crucial element for the formation of structure in the Universe.

Large, hydrodynamical simulations based on $\Lambda$CDM, such as Illustris \citep{Vogelsberger:2014b,Vogelsberger:2014,Genel:2014} and the Evolution and Assembly of GaLaxies and their Environments simulation \citep[EAGLE;][]{Schaye:2015,Crain:2015,McAlpine:2015,Schaller:2015b}, have had successes in recreating a universe with many characteristics similar to our own. These successes include the reproduction of the galaxy stellar mass function and its redshift evolution \citep[e.g.][]{Genel:2014,Furlong:2015}, the evolution of the mass-size relation of galaxies \citep[e.g.][]{Furlong:2017}, and local galaxy scaling relations \citep[such as the Tully-Fisher relation,][]{Tully:1977aa,Vogelsberger:2014,Ferrero:2017}. 

Whilst there is a general consensus on the need for dark matter in galaxy evolution theory, recent works have cast doubt on its relative dominance in galaxies in the distant past based on the shapes of their rotation curves. \citet{Lang:2017} used stacked H$\alpha$ emission from K-band Multi-Object Spectrograph (KMOS) observations to examine the average outer-kinematics of 101 star-forming galaxies with stellar masses $9.3 \lesssim \log M_{*}/M_{\odot} \lesssim 11.5$, over a redshift range $0.6 \lesssim z \lesssim 2.2$. The authors employed a novel method to construct position-velocity diagrams from the stacked flux that traces the average rotation curve of the galaxies out to $\sim$4 times the effective radius ($R_{\rm{e}}=4.6$ kpc, on average). They report this curve to exhibit a significant decline in its outer regions, seemingly at odds with the flat or rising rotation curve expected for local late-type galaxies of similar stellar mass. The authors conclude that the shape of the rotation curve in their analysis is consistent with that expected from a strongly baryon-dominated system, with a correspondingly small dark matter fraction and high levels of pressure support in its outer regions. 

In a partner study, \citet{Genzel:2017} present the individual rotation curves of six massive, star-forming galaxies at $ 0.9 \lesssim z \lesssim 2.4$, each of which exhibit a decline beyond their turnover radius. Like \citet{Lang:2017}, the authors conclude that their results imply massive galaxies at $z\approx$1--$2$ are both highly turbulent and strongly centrally baryon dominated with negligible dark matter fractions. These increased baryon fractions could arise in ``compaction'' scenarios \citep[e.g.][]{Dekel:2014,zolotov:2015} whereby baryons, that are more efficiently able to cool and condense than collisionless dark matter, fall to the centres of galaxy halos where they concentrate. This process may be facilitated via a number of potential mechanisms including extreme rates of gas accretion, an increased rate of mergers, or more secular scenarios involving gravitational disk instabilities -- each of which are more likely at earlier cosmic times around the peak of cosmic star-formation rate density.

These studies also raise the important question of whether or not such a result is a natural consequence of a $\Lambda$CDM universe, or whether there is some deviation from this long-standing cosmological consensus. Some recent studies find qualitative consistencies between the rotation curves presented in \citet{Lang:2017} and \citet{Genzel:2017} and those predicted for model galaxies with similar mass and at similar redshifts in $\Lambda$CDM cosmological, hydrodynamical simulations. \citet{Teklu:2017}, for example, select simulated galaxies at $z\approx2$, with stellar masses similar to those of the \citet{Genzel:2017} sample ($M_{*} > 5 \times 10^{10} M_{\odot}$) and with high cold gas mass fractions, from a 68 Mpc$^{3}$ volume within the Magneticum Pathfinder simulations. They find $\sim40$ percent of these model systems to exhibit significantly declining rotation curves (and with no signs of merger activity) similar to those presented in \citet{Genzel:2017}. \citeauthor{Teklu:2017} conclude that the declining curves are a result of significant pressure support in the disk of each model galaxy, as well as an abundance of baryonic matter in their centres that reduces the dark matter {\it fraction} in the central region (rather than a scarcity of dark matter). 

\begin{figure*}
\begin{minipage}[]{1.\textwidth}
\centering
\includegraphics[width=.85\textwidth,trim= 5 0 0 0,clip=True]{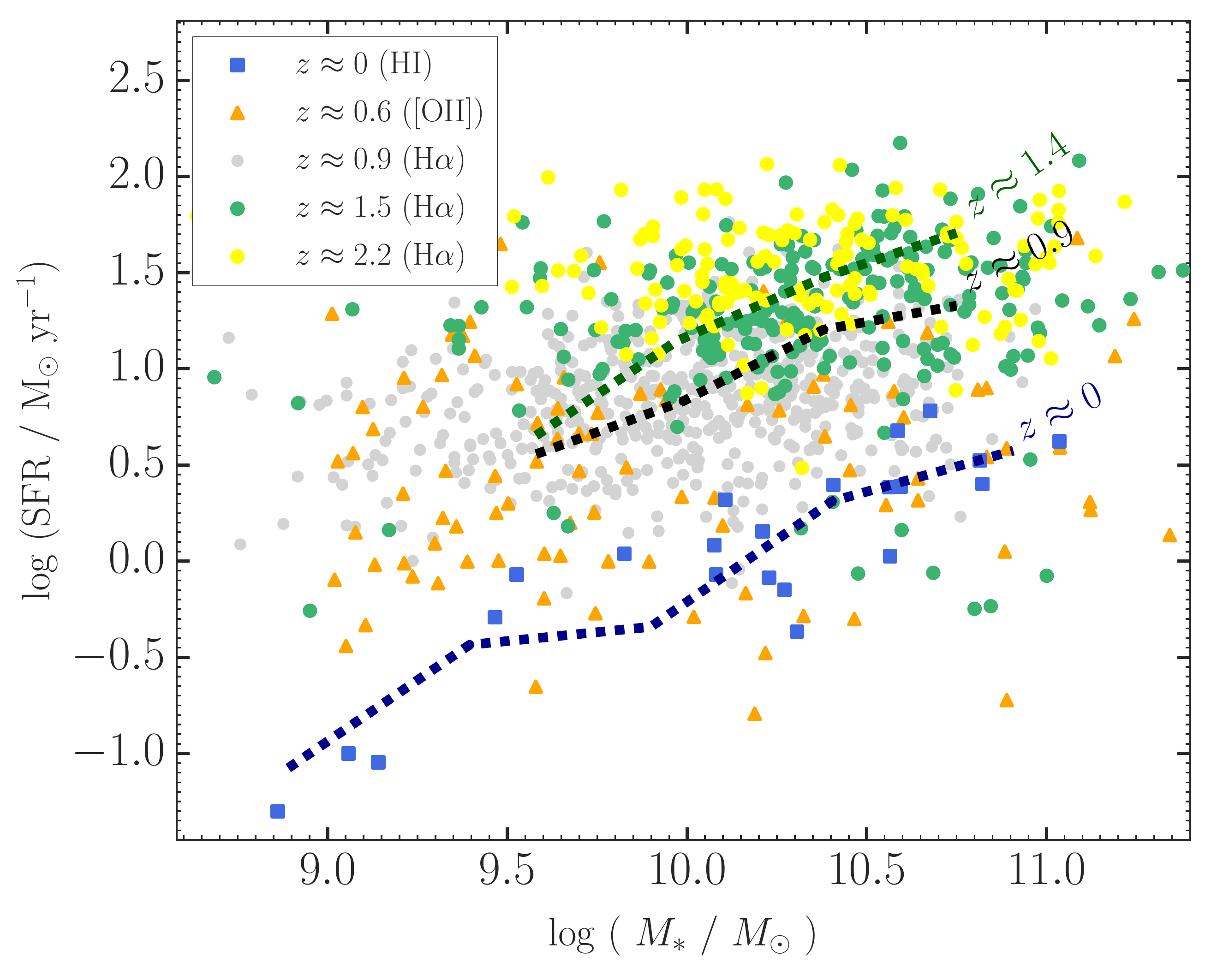}
\end{minipage}
\caption{%
Star-formation rate as a function of stellar mass for the THINGS ($z\approx0$), MUSE ($z\approx0.6$), KROSS ($z\approx0.9$), KGES ($z\approx1.5$), and publicly available KMOS$^{3\rm{D}}$ ($z\approx2.2$) samples. The dashed green and dashed black lines represent the median ``main sequence'' of star-forming galaxies at respectively $z\approx1.4$ and $z\approx0.9$ according to \citet{Karim:2011}. The blue dashed line represents the running median for star-forming, late-type (sersic indices $n\leq1.5$) SAMI Galaxy Survey \citep[e.g.][]{Bryant:2015} galaxies at $z\approx0$ \citep[][]{Johnson:2018,Tiley:2018}. The $z\approx0.6$ points have large scatter, likely the result of the more uncertain conversion between [O{\sc ii}] luminosity and star-formation in comparison to the corresponding conversion for H$\alpha$. In general the various galaxy samples of this work comprise ``normal'' star-forming galaxies with star-formation rates typical of their corresponding epoch. %
     }%
\label{fig:mainsequence}
\end{figure*}

The results of \citet{Lang:2017} and \citet{Genzel:2017} are potentially significant for galaxy evolution theory, bringing in to question the relative importance of dark matter at a key period for galaxy growth and evolution \citep[although see][]{Drew:2018}. The next step is therefore to expand the sample to include more galaxies, allowing to investigate trends with stellar mass and with redshift, as well as a comprehensive comparison between the observed shapes of galaxy rotation curves and those predicted from simulations. 

In this paper we exploit integral field spectroscopy observations of nebular emission from a sample of $\approx$1500 star-forming galaxies spanning $0.6 \lesssim z \lesssim 2.2$, along with observations of extended H{\sc i} emission from local galaxies, to measure the shape of galaxies' rotation curves over $\approx$10 Gyr of cosmic history. We compare the properties of the individual rotation curves of galaxies as well as combining the flux from various galaxy sub-samples to construct average rotation curves as a function of redshift, stellar mass, and central stellar mass surface density. To inform our interpretation, and as a means to interrogate $\Lambda$CDM  theory, we compare our results to model star-forming galaxies from the EAGLE simulation. 

In \S~\ref{sec:data} we describe the data used to construct our galaxy samples, including details of the observations and sample selection criteria of the constituent surveys from which they are drawn. In \S~\ref{sec:analysis} and \S~\ref{sec:buildingcurves} we detail our analysis methods, including the extraction of velocity maps, individual galaxy rotation curves, and subsequent stacked rotation curves. We also explore the biases inherent in different normalisation prescription used to construct the average rotation curves. Finally we describe the nebular flux stacking process used to construct average rotation curves that extend to larger radii than the observed curves of individual galaxies. In \S~\ref{sec:results}, we present the results of our analysis and an exploration of the shapes of galaxy rotation curves as a function of redshift and galaxy properties. We also include a thorough comparison of our results to trends for model galaxies in the EAGLE simulation. We provide concluding remarks in \S~\ref{sec:conclusions}.

\section{Observations \& Data}
\label{sec:data}

\begin{figure*}
\centering
\begin{minipage}[]{1\textwidth}
\centering
\includegraphics[width=1.\textwidth,trim= 118 10 170 20,clip=True]{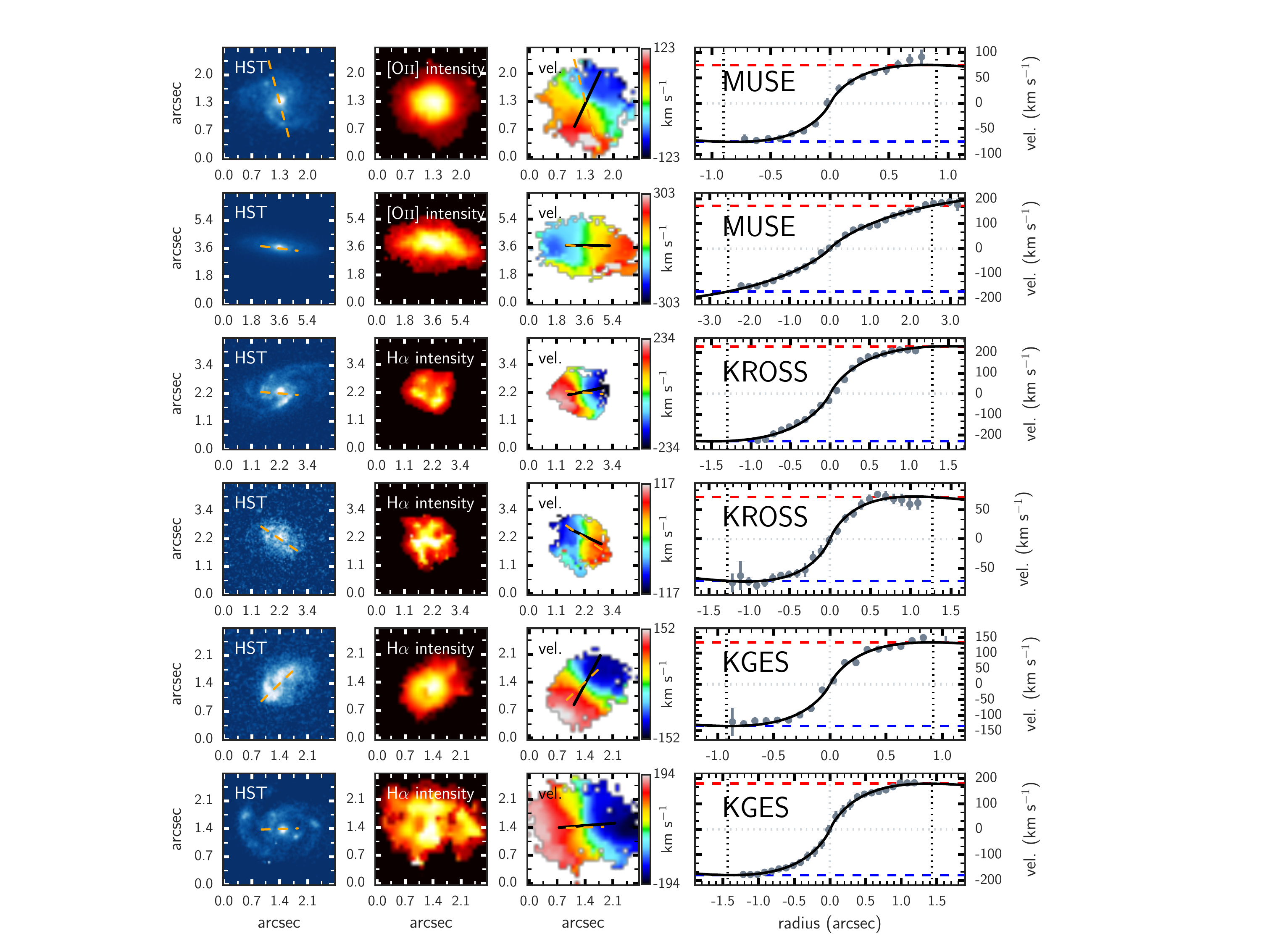}
\end{minipage}
\caption{%
Example data for galaxies in our sample. The panels from left to right are (1) The {\it Hubble Space Telescope} broadband image with the morphological position angle axis (measured from a two dimensional Gaussian fit to the image) overlaid as an orange dashed line. (2) The model H$\alpha$ intensity map constructed by integrating the best fit to the H$\alpha$ emission line in each spaxel. (3) The observed line-of-sight velocity map constructed by plotting in each spaxel the H$\alpha$ peak position from the simultaneous best fit to the H$\alpha$ and [N {\sc ii}] triplet -- this in velocity space and with respect to the galaxy systemic velocity. We overlay the kinematic position angle axis (corresponding to the maximum velocity gradient along a ``slit'' of width equal to half that of the PSF) as a black solid line. We also include the morphological position angle, again as an orange dashed line. (4) The observed rotation curve extracted from a ``slit'' of width equal to half that of the PSF, in spaxel wide steps. At each radius the observed rotation velocity is taken as the median across the slit. The black solid curve shows the best fit exponential disk model to the data. Black, dotted vertical lines are displayed at $\pm$3 times the disk scale radius ($R_{\text{d}}$). We show the observed velocity extracted at $\pm3R_{\text{d}}$ as the red and blue dashed line, respectively. %
     }%
\label{fig:panelkinematics}
\end{figure*}

We make use of observations of ionised and neutral gas emission from star-forming galaxies across a redshift range $0 \lesssim z \lesssim 2.2$. In this section we describe the data corresponding to each redshift. In \S~\ref{subsec:kmossamples} we describe the main KMOS samples selected from three large integral field spectroscopy surveys targeting H$\alpha$ and [N{\sc ii}] in star-forming galaxies covering a redshift range $0.6 \lesssim z \lesssim 2.2$. These include the KMOS Redshift One Spectroscopic Survey \citep[KROSS, $z\approx0.9$;][]{Stott:2016} and the KMOS Galaxy Evolution Survey (KGES, $z\approx1.5$; Tiley et al., in preparation) samples. To extend the redshift baseline of our sample to more distant epochs, and directly compare our results with previous similar studies, we also compare these with KMOS data for a sample of galaxies at $z\approx2.2$ from the European Southern Observatory (ESO) data archive comprising galaxies observed by the KMOS$^{3\rm{D}}$ Survey \citep{Wisnioski:2015}. We extend our analysis to lower redshift by including a sample of 96 galaxies observed in their [O{\sc ii}]$\lambda\lambda3726.2,3728.9$ emission with the Multi-Unit Spectroscopic Explorer \citep[MUSE;][]{Bacon:2010,Bacon:2015,Swinbank:2017} integral field unit, which have a median redshift of $z=0.67\pm0.01$. For a $z\approx0$ baseline we exploit H{\sc i} rotation curves for galaxies from The H{\sc i} Nearby Galaxy Survey \citep[THINGS;][]{Walter:2008}.

In Figure~\ref{fig:mainsequence}, we place each of the samples in context with one another on the star-formation rate-stellar mass plane, showing that each sample is comprised of galaxies that typically fall along the ``main sequence'' of star-formation at each epoch. In Figure~\ref{fig:panelkinematics}, we show example data for our sample, including broadband imaging, H$\alpha$ intensity maps, velocity maps, and rotation curves.

\subsection{KMOS Samples}
\label{subsec:kmossamples}

Here we describe the KMOS galaxy samples used in this work that comprise star-forming systems in three redshift slices, with median values ranging $0.9 \lesssim z \lesssim 2.2$, at the epoch of peak star-formation rate density in the Universe.

\subsubsection{KROSS}
\label{subsubsec:krossdata}

For galaxies at $z\approx0.9$ we exploit the KROSS sample. For descriptions of the KROSS sample selection and observations we refer the reader to \citet{Stott:2016} and \citet{Harrison:2017}. Briefly, KROSS comprises observations with KMOS of 795 galaxies at $0.6 \lesssim z \lesssim 1$. The observations target H$\alpha$, [N {\sc ii}]6548 and [N {\sc ii}]6583 emission from ionised gas that falls in the $YJ$-band ($\approx1.02$--$1.36 \mu$m). Target galaxies were selected to have $K_{\rm{AB}} < 22.5$ with priority (but not exclusivity) given to star-forming galaxies, as defined by a blue $(r-z)<1.5$ colour. Targets were selected in the well-studied extragalactic fields: the Extended {\it Chandra} Deep Field South (ECDFS), the Ultra Deep Survey (UDS), the COSMOlogical evolution Survey (COSMOS), and the Special Selected Area 22 field (SSA22). The ECDFS, COSMOS and parts of UDS all benefit from extensive {\it HST} coverage. 

KMOS \citep{Sharples:2013} consists of 24 individual integral field units (IFUs), each with a $2\farcs8 \times 2\farcs8$ square field-of-view (FOV), deployable in a $7'$ diameter circular FOV. The resolving power of KMOS in the $YJ$-band ranges from $R\approx3000$--$4000$. The KROSS observations were undertaken over two years, during ESO observing periods P92--P95\footnote{Programme IDs 092.B-0538, 093.B-0106, 094.B-0061, and 095.B-0035. The full sample also includes science verification data \citep[60.A-9460;][]{Sobral:2013,Stott:2014}}. The median seeing in the $YJ$-band for KROSS observations was $0\farcs7$. Reduced KMOS data results in a ``standard'' data cube for each target with $14 \times 14$ $0\farcs2$ spaxels. Each of these cubes is then re-sampled on to a spaxel scale of $0\farcs1$ during the data reduction process \citep{Stott:2016,Harrison:2017}.  

In this work we consider 551 KROSS galaxies with spatially resolved H$\alpha$ emission (following the selection described in \citealt{Harrison:2017}) and sufficient pixels in their velocity maps to measure a rotation velocity (\S~\ref{subsec:linemaps}). These galaxies have a median redshift of $z=0.85 \pm 0.04$, a median stellar mass of $10^{10.0 \pm 0.3} M_{\odot}$, and a median star-formation rate of $7 \pm 3\ M_{\odot}$ yr$^{-1}$ (where the uncertainty in each case is the median absolute deviation from the median itself).  

\subsubsection{KGES}
\label{subsubsec:kgesdata}

The $z\approx1.5$ galaxy sample is drawn from the KMOS Galaxy Evolution Survey (KGES), a recently completed 27 night GTO programme with KMOS. A detailed description of the KGES sample selection and observations will be presented in Tiley et al. (in preparation). In summary, KGES comprises KMOS observations of 285 galaxies at $1.3 \lesssim z \lesssim 1.5$ in COSMOS, CDFS, and UDS. The survey targets H$\alpha$, [N {\sc ii}]6548 and [N {\sc ii}]6583 from galaxy gas emission, redshifted into the $H$-band ($\approx1.46$--$1.85 \mu$m). Target galaxies were predominantly selected to be bright ($K<22.7$) and blue ($I-J<1.7$). The selection also favoured those systems with a previous H$\alpha$ detection, where available\footnote{including galaxies targeted as part of the FMOS-COSMOS Survey \citep{Silverman:2015,Kashino:2017}.}. In this work we consider 228 KGES galaxies detected in H$\alpha$ and with sufficient pixels in their velocity map to measure a rotation velocity (\S~\ref{subsec:linemaps}). The final KGES sample has a median redshift of $z=1.49\pm0.07$, a median stellar mass of $10^{10.3 \pm 0.3} M_{\odot}$, and a median star-formation rate of $21 \pm 10\ M_{\odot}$ yr$^{-1}$.  

The KGES observations were undertaken during ESO observing periods P95--P100\footnote{Programme IDs 095.A-0748, 096.A-0200, 097.A-0182, 098.A-0311, and 0100.A-0134.}. The resolving power of KMOS in the $H$-band ranges from $R\approx$3570--4555. The median seeing in the $H$-band for KGES observations was $0\farcs6$. 

\subsubsection{The KMOS$^{\ \it 3D}$ Survey}
\label{subsubsec:k3ddata}

Since we carry out similar analysis to \citet{Lang:2017}, it is prudent to test for systematics by examining galaxies within the same redshift interval as their work. We therefore construct a sample of $z\approx2.2$ star-forming galaxies from the KMOS$^{3\rm{D}}$ Survey. The data reduction process is similar to that for KROSS and KGES. For a full description of KMOS$^{3\rm{D}}$ see \citet{Wisnioski:2015}. 

In this work we exploit 145 KMOS$^{3\rm{D}}$ galaxies that fall in the upper redshift slice of the survey (spanning $1.9 \lesssim z \lesssim 2.7$), are resolved in H$\alpha$, and with sufficient pixels in their velocity maps to measure a rotation velocity (\S~\ref{subsec:linemaps}). This sample has a median redshift of $z=2.3 \pm 0.1$. The $K$-band KMOS observations of these systems target the H$\alpha$, [N {\sc ii}]6548 and [N {\sc ii}]6583 emission from these galaxies. The median stellar mass of the targets is $10^{10.3 \pm 0.3} M_{\odot}$, and the median star-formation rate is $32 \pm 14\ M_{\odot}$ yr$^{-1}$. 

\subsection{Lower Redshift Comparison Samples}
\label{subsec:compsamples}

In this section we describe two lower-redshift comparison samples of galaxies that we construct in order to inform and extend our interpretation of the KMOS samples. 

\subsubsection{THINGS}
\label{subsubsec:thingsdata}

For a comparison sample of galaxies in the local Universe we exploit extended H{\sc i}-derived rotation curves for $z\approx0$ galaxies from THINGS, which obtained high-quality observations of the extended H{\sc i} emission for 33 nearby galaxies encompassing a wide range of galaxy morphologies, star-formation rates, luminosities and metallicities. For this work, we consider 22 star-forming galaxies from THINGS, with $M_{*} \gtrsim 10^{9} M_{\odot}$ and star-formation rates $\gtrsim 0.05\ M_{\odot}$ yr$^{-1}$, for comparison with our higher-redshift star-forming samples.

We note here the existence of the {\it Spitzer} Photometry and Accurate Rotation Curves (SPARC) sample that offers publicly available high-quality H{\sc i} and H$\alpha$ extended rotation curves for $175$ nearby galaxies. In this work, for convenience, we prefer to exploit THINGS given that tabulated values of key galaxy properties are readily available. However, we stress that our results are robust to our choice of $z\approx0$ baseline sample with the shapes of the average THINGS and SPARC rotation curves (within $6R_{\rm{d}}$, where $R_{\rm{d}}$ is the disk-scale radius; \S~\ref{sec:analysis}) in excellent agreement. 

\subsubsection{MUSE}
\label{subsubsec:musedata}

For an intermediate-redshift sample, we use the star-forming galaxies from \citet{Swinbank:2017}, drawn from MUSE observations of 17 extragalactic fields, including those taken between February 2014 and February 2015 as part of the commissioning and science verification stage of the instrument's construction \citep[see e.g.][]{Richard:2015,Husband:2015,Contini:2016}. For detailed descriptions of the MUSE sample selection and observations see \citet{Swinbank:2017}. Briefly, our MUSE sample comprises observations of 431 galaxies with spatially resolved [O{\sc ii}] emission that falls within the wavelength coverage of MUSE ($4777$--$9300$\AA), corresponding to a redshift range of $0.3 \lesssim z \lesssim 1.5$. For this work we select a sub-sample of 96 of these systems that are spatially resolved in [O{\sc ii}], with redshifts $z \leq 0.8$, stellar masses $M_{*} \gtrsim 10^{9} M_{\odot}$ (to match the effective mass cut of the KROSS galaxies), and sufficient pixels in their velocity maps to measure their rotation (\S~\ref{subsec:linemaps}). The resultant sample has a median redshift, stellar mass, and star-formation rate of respectively $z=0.67 \pm 0.09$, $10^{9.8 \pm 0.5} M_{\odot}$, and $3 \pm 2\ M_{\odot}$ yr$^{-1}$. 

\section{Analysis}
\label{sec:analysis}

The goal of this work is to measure the shape of the rotation curves of typical star-forming galaxies out to large radii as a function of redshift, stellar mass, and stellar mass density. Since the shape of a galaxy's rotation curve should be intimately linked to its mass distribution, and hence its dark matter content, we aim to infer to what extent the dark matter fraction of galaxies has evolved over cosmic time and explore which processes or mechanisms may be driving any such changes. 

The challenge for this work lies in the difficulty in tracing the rotation curves of higher-redshift galaxies out to sufficiently large distances as to begin to reliably probe the contribution of the dark halo. The rotation curves of local galaxies have been well constrained out to large radii. To trace galaxies' kinematics out to 10's of kpc these studies have primarily used bright emission lines from spatially extended gas in galaxies or atomic hydrogen (H{\sc i}) 21 cm emission lines. Unfortunately, such an approach is not currently possible for more distant galaxies; H{\sc i} is much more difficult to detect with increasing redshift -- routine detections of 21cm emission from galaxies up to $z\approx0.8$, for example, will require the complete Square Kilometre Array \citep[SKA,][]{Abdalla:2015,Yahya:2015}. And rotation curves at $z \gtrsim 1$ in H{\sc i} may only be measured with the capabilities of the still-hypothetical SKA2. 

To reliably trace galaxy kinematics at higher-redshift, observations, particularly those of a spatially-resolved, integral field spectroscopy nature, rely on bright nebular emission lines. The most widely used tracer is H$\alpha$ (along with the [N{\sc ii}]6548,6583 doublet) that traces the warm ionised gas surrounding young stars in galaxies. Several studies have measured the dynamics of the ionised gas in galaxies at $z\approx1$--$3$ using near-infrared integral field spectroscopy \citep[e.g.][]{ForsterSchreiber:2009,Davies:2015,Wisnioski:2015,Stott:2016,Teodoro:2016,Beifiori:2017,Turner:2017}. However, the integration times adopted by these surveys typically only allow the rotation curves of individual galaxies to be traced out to a few times the galaxy disk-scale radius ($\approx3R_{\rm{d}}$, equivalent to $\approx1.8$ times the half-light radius, $R_{\rm{h}}$ or $\approx7$ kpc for a $\approx10^{10}$ $M_{\odot}$ galaxy at $z\approx0.9$). This is insufficient to repeat the experiments of Rubin et al. at high redshift and robustly probe the outer regions of a galaxy's rotation curve, and provides little diagnostic use in determining total galaxy dark matter fractions. 

To compare to $z\approx0$, in this work we require measurements of the shape of galaxy rotation curves out to $\approx6R_{\rm{d}}$ for galaxies at $0.6 \lesssim z \lesssim 2.2$. This will match the typical measurements of galaxy rotation curves in the local Universe \citep[e.g.][]{Catinella:2006,Catinella:2007} and facilitate direct comparisons between the epochs. As will be discussed later, this is also sufficient radial extent to robustly measure the outer slope of the curves. However, the depths of our integral field spectroscopy observations mean we only detect sufficient nebular emission to trace the galaxy rotation curve out to $\pm 6R_{\rm{d}}$ for $6$ percent of individual systems in our combined samples - and with the majority of these galaxies in a single redshift bin ($z\approx0.9$). We must therefore sacrifice the detail that one gains in considering the rotation curves of individual galaxies, and instead combine the signal from many galaxies to construct average curves. 

In this section we describe our measurements of the properties of individual galaxies that are required as precursors to the construction of average galaxy rotation curves. In \S~\ref{subsec:stellarmass} we provide details on the galaxy stellar masses and star-formation rates. In \S~\ref{subsec:imageandalign}, we outline methods used to spatially align our KMOS cubes with the available broadband imaging for each galaxy in our sample, as well as extracting measurements of galaxy sizes from the same imaging. In \S~\ref{subsec:linemaps} we detail the construction of kinematic maps from the data cube of each galaxy in our sample, and the subsequent extraction of rotation curves for each individual galaxy. A detailed description and exploration of the methods used to construct the average rotation curves for galaxies in our sample is presented in \S~\ref{sec:buildingcurves}.

\subsection{Stellar Masses and Star Formation Rates}
\label{subsec:stellarmass}

Stellar masses for the MUSE, KROSS, KGES, and KMOS$^{3\rm{D}}$ samples were calculated via comparison of suites of model spectral energy distributions to broadband photometry for each galaxy typically spanning the visible and near-infrared bandpasses, adopting a Chabrier \citep{Chabrier:2003} initial mass function and allowing for a range of star formation histories, metallicities and dust extinction \citep{Santini:2015,Swinbank:2017,Harrison:2017}. Stellar masses for THINGS galaxies were calculated via conversion from their infrared (3.6$\mu$m) flux \citep{deBlok:2008,Querejeta:2015}. Where the assumed initial mass function differs from Chabrier, we convert them appropriately for this work\footnote{We note that one may expect stellar mass estimates derived via different methods or spectral energy distribution fitting codes to typically deviate by $\pm0.2$ dex \citep{Mobasher:2015}. We have verified that the results presented in this work are robust to a systematic change in stellar mass of $\pm0.2$ dex.}.

Star formation rates for galaxies in our sample observed with KMOS are calculated via conversion from their extinction-corrected H$\alpha$ luminosities in the manner of \citet{Harrison:2017}. Similarly, star-formation rates for galaxies observed with MUSE are calculated via conversion of their extinction corrected [O{\sc ii}] flux according to the prescription of \citet{Kewley:2004}. Star-formation rates for THINGS galaxies are sourced from \citet{Walter:2008}, calculated from published flux values \citep{Leroy:2008}.

\subsection{Cube Alignment and Stellar Sizes}
\label{subsec:imageandalign}

\subsubsection{Cube Alignment}

To ensure we consider the same physical regions of galaxies in both the broadband imaging and the integral field spectroscopy data cubes, we spatially align each data cube in our sample with the highest-quality broadband image. We construct a continuum image from each cube using the median flux of each spectrum of each spaxel, after masking any line emission in the cube and performing a $2$-$\sigma$ clip to the spectrum to exclude significant sky residuals. This map is then astrometrically aligned with the broadband image according to the position of the peak of the continuum (see \citealt{Harrison:2017} for a full description).

\subsubsection{Stellar sizes}
\label{subsubsec:stellarsizes}

For the size of each galaxy we adopt the stellar disk-scale radius $R_{\rm{d}}=0.59R_{\rm{h}}$, where $R_{\rm{h}}$ is the projected stellar half-light radius, as measured from the highest resolution and longest wavelength (optical or near-infrared) broadband imaging available for each galaxy sub-sample. For each galaxy in KROSS and KGES we perform a two-dimensional Gaussian profile fit to the broadband image to recover the position angle of the galaxy's morphological axis, and its axial ratio. We then construct a curve-of-growth by summing the image flux within elliptical annuli matched in orientation and axial ratio to the galaxy, and incrementally increasing in size. We measure $R_{\rm{h}}$ as the semi-major axis of the ellipse containing 50 percent of the maximum of the summed flux in the image. We inspect each curve-of-growth to ensure it asymptotes to a maximum value. To test the validity of this method, we compare our measure of the half-light radius of KGES galaxies to those of \citet{vanderWel:2012} as measured from detailed S\'ersic model fits to the $H$-band {\it HST} imaging in the Cosmic Assembly Near-Infrared Deep Extragalactic Legacy Survey \citep[CANDELS;][]{Grogin:2011}. For those 127 (out of 285) KGES galaxies in our sample that overlap with galaxies examined by \citet{vanderWel:2012}, we find excellent agreement between the two measures with a median difference of $0\farcs0 \pm 0\farcs1$. 

For the stellar sizes of KMOS$^{3\rm{D}}$ galaxies, we adopt the effective radius measurements of \citet{vanderWel:2012} that use Sersic fits to the $H$-band {\it HST} image for each galaxy from CANDELS. We convert these to disk-scale radii with the same scaling factor as the other samples.

For a measure of stellar size for the THINGS galaxies we adopt the {\it k\_r\_eff} $K_{s}$-band half-light radius from the 2MASS Extended Source Catalog. These radii are converted to disk-scale radii in the same manner as for the other samples.

The stellar sizes for the MUSE galaxies were measured by \citet{Swinbank:2017}. For $\approx60$ percent of the total sample, sizes were measured from {\it HST} images, fitting a two-dimensional Sersic profile to define the galaxy centre and position angle and then constructing a curve of growth with ellipses of increasing size, matched in ellipticity and position angle to the initial fit. For the remainder of the sample (those without {\it HST} imaging), sizes were measured directly from the MUSE continuum maps, deconvolving for the PSF. Once again, half-light radii are converted to disk-scale radii using the same scaling factor as for the KGES and KROSS samples.

\subsection{Emission Line Maps and Rotation Curves}
\label{subsec:linemaps}

H$\alpha$ imaging and kinematic maps were extracted from the galaxy data cubes in the manner of \citet{Stott:2016}. Maps were extracted from the KMOS cubes via a simultaneous triple Gaussian profile fit to the H$\alpha$, [N {\sc ii}]6548 and [N {\sc ii}]6583 emission lines in each (continuum-subtracted) spectrum of each spaxel for each cube. The central velocity and width of the H$\alpha$ and [{N{\sc ii}] lines are coupled during the fit. The velocity dispersions are deconvolved for the instrumental resolution. If the H$\alpha$ $S/N < 5$ for a given spaxel, a larger area of  $3 \times 3$ spaxels was considered, and $5 \times 5$ spaxels, as required. If at this point the $S/N$ was still less than 5, that spaxel is excluded from the final maps. 

Similarly, maps were extracted from the MUSE cubes following the same method but instead performing a double Gaussian profile fit to the [O{\sc ii}] emission line doublet in each spaxel of each cube. The widths of each of the emission lines in the doublet are coupled (and deconvolved to account for instrumental broadening), and the same adaptive binning process is employed during the fitting.

To measure the maximum (observed) rotation velocity of each galaxy we extract a rotation curve along the major kinematic axis of each. We measure this axis and the position angle for each of the KMOS and MUSE galaxies by rotating their line-of-sight velocity maps in 1$^{\circ}$ steps about the continuum centre. For each step we calculate the velocity gradient along a horizontal ``slit'', centred on the continuum peak and with width equal to half the full width at half maximum (FWHM) of the point spread function (PSF). We select as the position angle the choice that maximises the velocity gradient. We extract the velocity and its uncertainty along the major kinematic axis as respectively the median and median absolute deviation velocity along the pixels perpendicular to the ``slit'' at each step. Example rotation curves derived in this manner are shown in Figure~\ref{fig:panelkinematics}. We note that whilst these rotation curves are derived from the galaxy velocity maps, they agree well with position-velocity diagrams extracted from the individual integral field spectroscopy cube for each galaxy (see \S~\ref{subsec:specs}).

We also note that in this work we prefer this direct method to extract rotation curves in comparison to a more complex two- or three-dimensional, forward-modelling analysis of the galaxy kinematics \citep[e.g.][]{Stott:2016,Tiley:2016,Teodoro:2016} since it relies on the least number of assumptions with regards to the physical properties of the galaxies in our samples. Of course, as a result of this simplicity, the rotation curves initially derived via our direct method will suffer more from the effects of beam smearing than curves derived from a forward-modelling approach. However, these affects are addressed and mitigated (down to a $\approx10$ percent level) at the point at which we normalise the rotation curves in rotation velocity and radius (see \S~\ref{subsec:stackRC}). Importantly, this correction takes place before we interpret the shapes of the curves.

\section{Constructing Average Galaxy Rotation Curves}
\label{sec:buildingcurves}

To construct average rotation curves for galaxies we explore two broad methods: Section~\ref{subsec:stackRC} presents the properties of the median stacks of the {\it individual galaxy rotation curves}, where each curve is first normalised in size and velocity. The choice of values by which to normalise the curves is not an obvious one and is therefore discussed in detail. In \S~\ref{subsec:specs}, we derive average normalised rotation curves from position-velocity diagrams constructed from the stacked galaxy {\it emission}, allowing us to extend the average curve out to larger scale radii than for the median stacks of the individual rotation curves. We describe the construction of the stack, and the subsequent extraction of the normalised curves. We then present the resultant curves as a function of redshift and key galaxy properties.

\subsection{Stacking Individual Rotation Curves}
\label{subsec:stackRC}

We begin by constructing simple median stacks of the {\it individual galaxy rotation curves} in bins of redshift. To do this we must first normalise each galaxy's rotation curve in both size and velocity. However, the choice of values by which to normalise the curves is not obvious (since rotation curves do not represent a linear property of galaxies, characterised by a single characteristic scale or quantity). As such this simple analysis provides a convenient means with which to test the effect of different scaling prescriptions on the properties of the final average rotation curve. We therefore stack the curves normalised in two ways. First, we normalise each curve in radius by sampling it at multiples of the stellar light disk-scale radius ($R_{\rm{d}}$, as defined in \S~\ref{subsec:imageandalign}), accounting for the point-spread-function (PSF) beam smearing by adding in quadrature with the width of the best fit Gaussian to the PSF ($\sigma_{\rm{psf}}$). The curve is sampled at $\rm{N}R_{\rm{d}}^{\prime} \equiv \sqrt{(\rm{N} \times R_{\rm{d}})^{2} + \sigma_{\rm{psf}}^{2}}$, where $\rm{N}$ is a set of constants (e.g. $\rm{N} = -6, -5, -4,$ ...$, 6$). We normalise in velocity by the {\it observed} velocity at $3R_{\rm{d}}^{\prime}$ ($v_{3R_{\rm{d}}^{\prime}}$), that should sample the galaxy rotation curve beyond its turnover. For simplicity we refer to these curves as {\it stellar-scaled}. 

\begin{figure*}
\centering
\begin{minipage}[]{1\textwidth}
\centering
\includegraphics[width=.95\textwidth,trim= 20 10 20 10,clip=True]{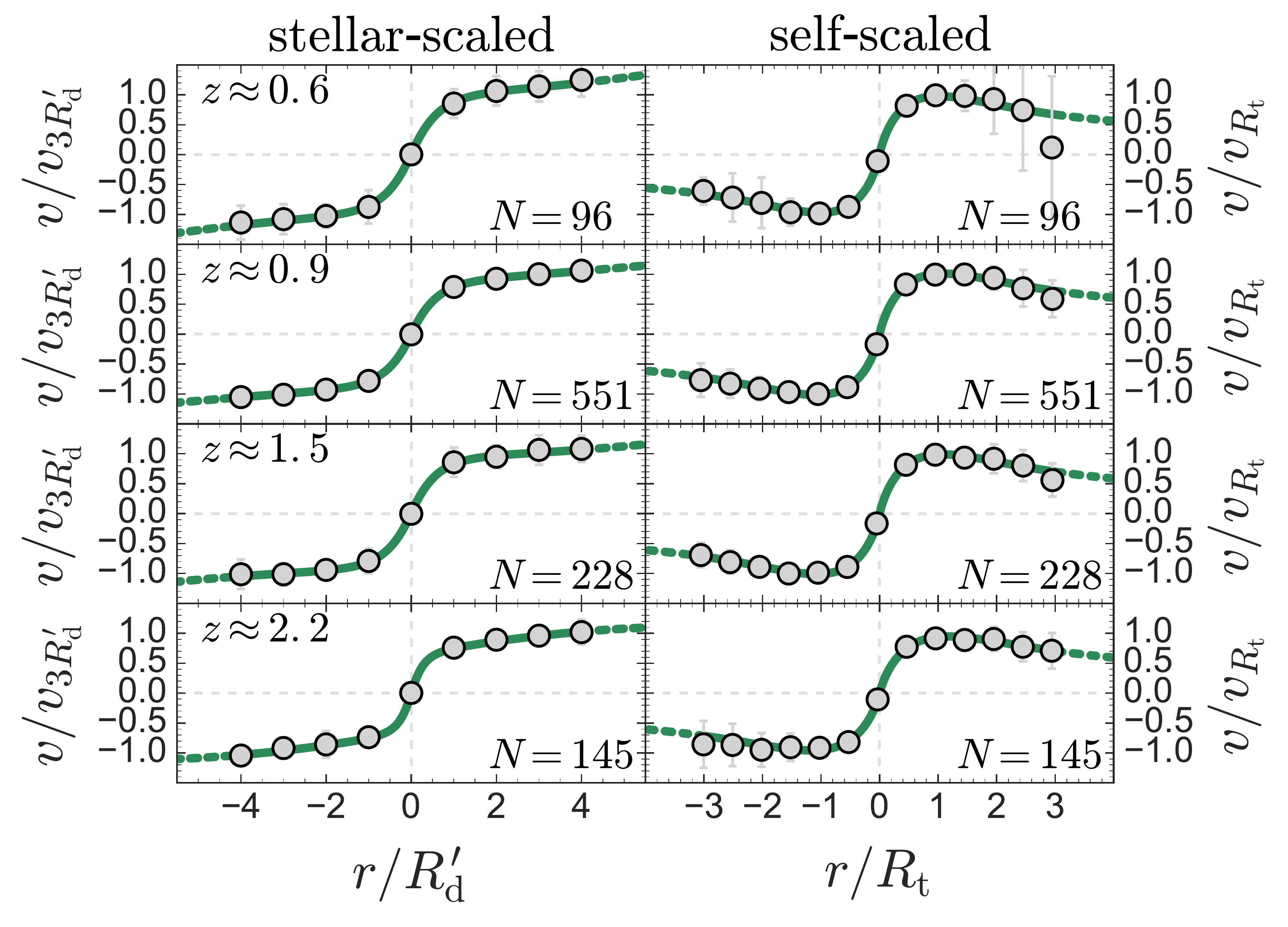}
\end{minipage}
\caption{%
Median stacks of the rotation curves extracted individually from each galaxy in the $z\approx0.6$, $z\approx0.9$, $z\approx1.5$, and $z\approx2.2$ samples. {\it For the same data}, we employ two normalisation prescriptions. {\bf Left column:} Median stack normalised by the modified stellar disk-scale radius ($R_{\rm{d}}^{\prime}$) and the velocity at $3R_{\rm{d}}^{\prime}$ ($v_{3R_{\rm{d}}^{\prime}}$). {\bf Right column:} Median stack normalised by the galaxy turnover radius ($R_{\rm{t}}$ i.e. the rotation curve inflection point) and the velocity at $R_{\rm{t}}$ ($v_{R_{\rm{t}}}$). For each median rotation curve (grey points), we also plot the best fit exponential disk plus dark halo model (solid green line; the dashed green line represents extrapolation of the same model beyond the extent of the data). The shape of the median average rotation curve {\it for the same galaxies} starkly differs depending on the normalisation technique.
     }%
\label{fig:medRCstacks}
\end{figure*}

\begin{figure*}
\centering
\begin{minipage}[]{1.\textwidth}
\centering
\includegraphics[width=.95\textwidth,trim= 20 0 10 10,clip=True]{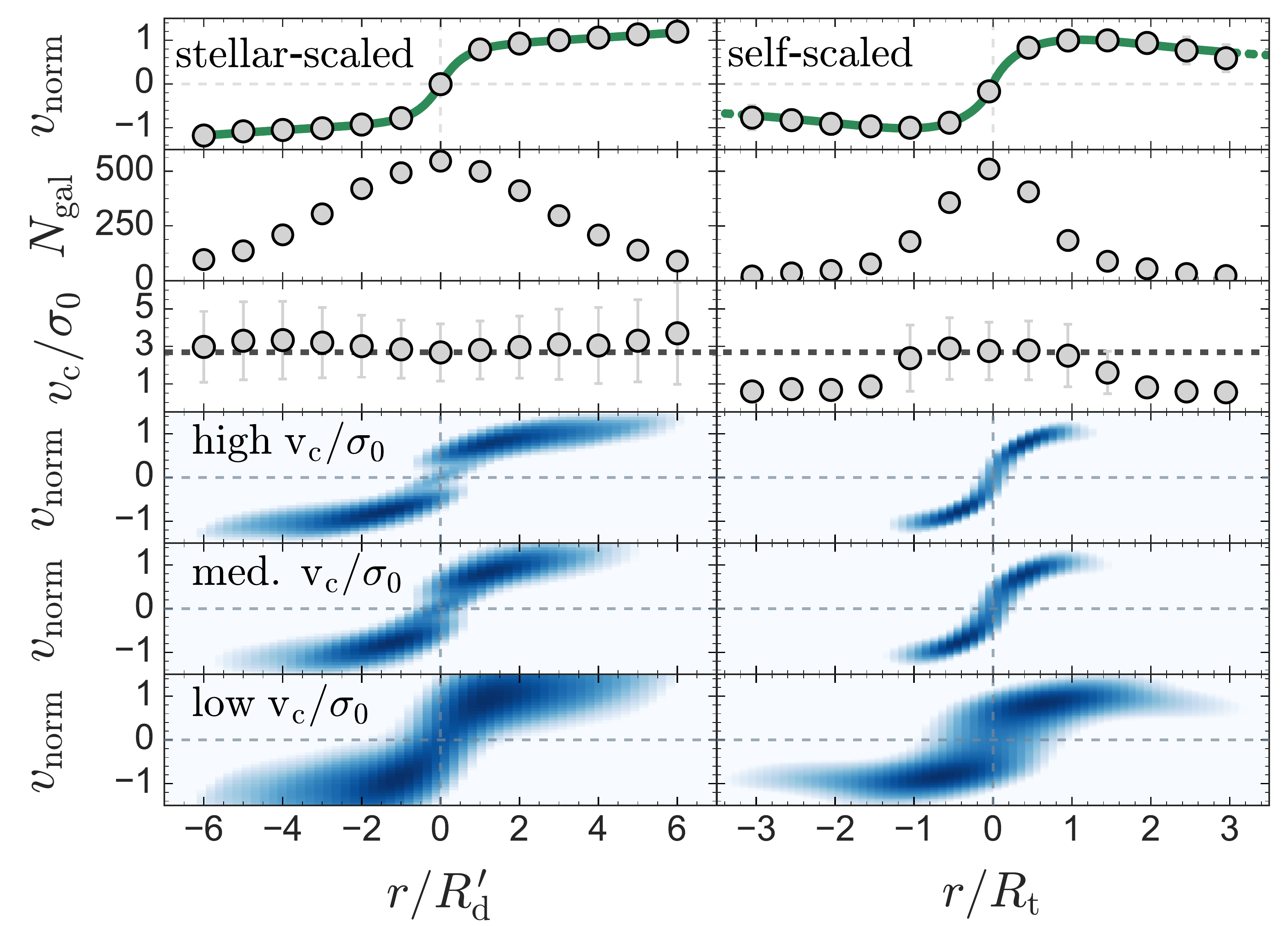}
\end{minipage}
\caption{%
An exploration of the inherent biases associated with the stellar-scaled (left column) and self-scaled (right column) rotation curve normalisation prescriptions, using the KROSS $z\approx0.9$ sample as an illustration. Starting from the top row downward, {\bf Top Row:} The median stacked rotation curves (grey points) with the best fit exponential disk plus dark halo model (solid green line -- dashed green line when extrapolated beyond the data). {\bf Second Row:} The corresponding number of contributing galaxies to the median curve at each radius.  {\bf Third Row:} The median (beam smearing corrected) rotation-to-dispersion ratio of galaxies contributing to the rotation curve at each radius. The dashed blacked line represents the median value for the total sample. {\bf Lowest Three Rows:}  The two-dimensional density distribution of the individual rotation curves, as measured by the log of the Gaussian kernel density estimate of rotation curve points (light blue to dark blue for low to high density). The distributions are shown for three different bins in (beam smearing corrected) intrinsic rotation-to-dispersion ratio (high: median $v_{\rm{c}}/\sigma_{0} = 5.8\pm0.3$, medium: $v_{\rm{c}}/\sigma_{0} = 2.69\pm0.08$, low: $v_{\rm{c}}/\sigma_{0} = 0.97\pm0.06$). For both normalisation prescriptions we see a reduction in the number of galaxies effectively contributing to the median curve as a function of increasing multiples of the scale radius. However, this decline occurs more rapidly for the self-scaled curve compared to the stellar-scaled curve. For the stellar-scaled rotation curve, the median $v_{\rm{c}}/\sigma_{0}$ of the  galaxies contributing to the curve remains approximately constant with increasing scale radius. However, the self-scaled rotation curve is biased toward low $v_{\rm{c}}/\sigma_{0}$ systems at larger radii. From the lowest three rows it is clear that this bias toward low $v_{\rm{c}}/\sigma_{0}$ systems with increasing radius also drives the shape of the self-scaled curve; only the lowest $v_{\rm{c}}/\sigma_{0}$ bin contains individual self-scaled rotation curves that significantly extend beyond $\pm1R_{\rm{t}}$. The shape of the self-scaled rotation curve beyond $\pm1R_{\rm{t}}$ is therefore entirely dictated by a minority of galaxies (only 28 percent of the KROSS galaxies in the stack fall within the lowest $v_{\rm{c}}/\sigma_{0}$ bin) with very low ratios of rotational-to-dispersive internal motions. %
     }%
\label{fig:biascheck}
\end{figure*}

To test the extent to which beam smearing effects impact the shape of the average stellar-scaled curves, we perform the following tests. For a KROSS-like ($z\approx1$) distribution in the ratio of the galaxy size to the size of the PSF, we generate beam smeared iterations of the velocity map for a baryonic exponential disk plus dark matter halo model where the rotation curve shape is matched to the average for model EAGLE galaxies at $z\approx1$ (see \S~\ref{subsubsec:eaglerc}). Extracting the stellar-scaled curve from each iteration we find the shape of the median average curve to be accurate to within $\approx10$ percent with respect to the rotation curve extracted from the intrinsic, non-beam smeared model and scaled in radius and velocity by $R_{\rm{d}}$ and the velocity at $3R_{\rm{d}}$, respectively. In Appendix~\ref{subsec:seeingeffect}, using the KROSS sample as a test case, we also show that the shape of the median average of the individual stellar-scaled rotation curves is insensitive to cuts in the ratio of the galaxy size to PSF size. Furthermore, in Appendix~\ref{subsec:altfittingeffect} we show, using the KGES ($z \approx 1.5$) galaxies as an example, that the shape of the average stellar-rotation curve does not depend on whether we fit the nebular emission lines in each spaxel of each data cube using a Gaussian triplet model (as described in \S~\ref{subsec:linemaps}) or instead a Gaussian-hermite triplet model (that should better account for any asymmetric deviation of the emission line shapes from Gaussian as a result of beam smearing effects). We therefore conclude that the shape of the average stellar-scaled rotation curve is robust to the effects of beam smearing in the observational data and has a shape that is an accurate representation of that of the {\it intrinsic} average (scaled) rotation curve for the galaxies used to construct it.

In addition to the stellar-scaled curves, to mimic the analysis of \citet{Lang:2017}, we also normalise each rotation curve by its dynamical turnover radius ($R_{\rm{t}}$), and the velocity at this turnover radius ($v_{R_{\rm{t}}}$) as measured from the best fit exponential disk model to the rotation curve. The model velocity ($v$) as a function of radius ($r$) takes the form

\begin{equation} \label{eq:mod} 
(v(r)-v_{\rm{off}})^{2}=\frac{(r-r_{\rm{off}})^2 \pi G \mu_{0}}{h}(I_{0}K_{0}-I_{1}K_{1})\,\,,
\end{equation}

\noindent where $G$ is the gravitational constant, $\mu_{0}$ is the peak mass surface density, $h$ is the disk scale radius, and $I_{\rm{n}}K_{\rm{n}}$ are Bessel functions evaluated at $0.5r/h$. We include parameters to allow for a global offset of the rotation curve in both velocity and radius space; $v_{\rm{off}}$ is the velocity at $r=0$, $r_{\rm{off}}$ is the radius at which $v=0$. Each rotation curve is corrected for non-zero values of either $v_{\rm{off}}$ or $r_{\rm{off}}$ before it is considered for further analysis. Similarly, once the observed curve is corrected, we set $v_{\rm{off}}=r_{\rm{off}}=0$ in the model. The velocity at the turnover radius, $v_{R_{\rm{t}}}$, is equivalent to the velocity at $2.2h$ or the maximum velocity of the model. Since for this method we are normalising by a radius measured from the curves themselves, we call these curves {\it self-scaled}. 

For each of the two choices of scaling (stellar-scaled, and self-scaled) we then linearly interpolate all of the individual curves on to a common radial axis. To construct a median rotation curve in each case we measure the median velocity of the re-sampled curves as a function of radius. As a measure of uncertainty we adopt the median absolute deviation of the rotation curves (with respect to the median value itself). We note that if an individual galaxy rotation curve is missing data at a given (scale) radius then by necessity it does not contribute to the average curve at that point. Therefore we expect a different number of galaxies to contribute to the average curve at different radii, and for this number to generally decrease with increasing radius. This is discussed in further detail in \S~\ref{subsubsec:normbiases}.

The resultant median rotation curves are presented in Figure~\ref{fig:medRCstacks}. We plot the curves out to radii for which the error in the velocity is less than 50 percent of the peak for both the positive and negative halves of the curve for the majority of the redshift bins. We note that the stellar-scaled median rotation curve for the KROSS galaxies extends beyond $4R_{\rm{d}}^{\prime}$ but that at these radii its shape is dictated by an increasingly small number of galaxies. Since the average rotation curve in every other redshift bin extends to only $\approx4R_{\rm{d}}^{\prime}$, we truncate the average KROSS curve accordingly (see Figure~\ref{fig:biascheck} for the extended median average KROSS curve). The shapes of the median average rotation curves in Figure~\ref{fig:medRCstacks} strongly depend on the normalisation scheme applied to the individual curves used to construct them. If they are self-scaled in the manner of \citet{Lang:2017}, the resultant average curves exhibit strong declines beyond their peaks for the radii probed by the data (agreeing with the steep declines for galaxy rotation curves at $z\approx1$--$2$ measured by \citeauthor{Lang:2017}). If instead {\it for the same samples of galaxies} the curves are normalised by the scale of the stellar light (stellar-scaled), they remain flat or continue to rise out to the maximum radii probed by the data. This is the case in each of the four redshift bins.

We note here that the individual galaxy rotation curves are over-sampled, with the spatial sampling ($0\farcs1$ steps) smaller than the typical seeing ($\sim0\farcs5$--$0\farcs7$) of the observations. However, in Appendix~\ref{subsec:binningeffect} we show that the average rotation curves (both stellar-scaled and self-scaled) are robust to this choice of spatial sampling, demonstrating that their shapes do not significantly change when the sampling size is increased by a factor of four (making each spatial bin in the rotation curve nearly independent).  

\subsubsection{Investigating Biases in Normalisation Prescriptions}
\label{subsubsec:normbiases}

The shape of the average normalised rotation curve for galaxies differs as a function of the scaling prescription employed to construct it. This is clearly a concern with respect to their interpretation in the context of galaxy evolution; subtle differences in how the individual curves are renormalised can result in significant differences in how the final, average rotation curve may be interpreted. 

At first inspection, there are no strong reasons to prefer one scaling prescription over the other, except that intuitively one might expect that renormalising by galaxies’ rotation curve turn-over radius would give more weight to any central bulge contribution (that is more Keplerian in its dynamics), if it is present. In contrast, renormalising instead by the spatial extent of the stars provides a radial scale that is completely independent of the rotation curve (although still influenced by a dominant bulge contribution). In this case, one might expect the average rotation curve to more equally reflect both the bulge and disk components (if present to the extent that they significantly affect the shape of the rotation curve). Similarly, if the observed rotation curve remains flat or continues to rise within the extent of the data, one might then expect that the turnover radius $R_{\rm{t}}$ of the best fitting exponential disk model will preferentially tend toward a value close to the maximum radius probed by the data to improve the goodness of the fit. Normalising in radius by $R_{\rm{t}}$ in these cases would thus compress the entire rotation curve to fall within $\pm1R_{\rm{t}}$. The stellar scaled curves, however, should not suffer from this effect since $R_{\rm{d}}^{\prime}$ is not directly dependent on the shape of the curve itself. 

In Figure~\ref{fig:biascheck}, we quantitatively test this conjecture on the merits of each scaling prescription by examining in detail the biases inherent in each. For this test, we use the KROSS $z\approx0.9$ galaxy sample as an example. We plot both the median stellar-scaled and self-scaled rotation curves for the KROSS sample, along with the number of galaxies effectively contributing at each radius. We also show the median rotation-to-dispersion ratio $v_{\rm{c}}/\sigma_{0}$, where $v_{\rm{c}}$ and $\sigma_{0}$ are respectively the {\it intrinsic} circular velocity and the {\it intrinsic} velocity dispersion measurements made in \citealt{Harrison:2017} and \citealt{Johnson:2018} for those galaxies. We stress that both $v_{\rm{c}}$ and $\sigma_{0}$ are corrected for the effects of beam smearing, according to the methods described in \citealt{Johnson:2018}. Figure~\ref{fig:biascheck} shows that for both the stellar-scaled and self-scaled curves the number of galaxies contributing to the average curve declines with increasing (scale) radius. In other words, more galaxies contribute to the inner parts of the rotation curves (for which the majority of galaxies have sufficiently nebular flux to sample) than the outer parts (for which only those systems with the brightest and most spatially-extended nebular emission will be able to contribute). However, this decline is much more rapid for the self-scaled curve than for the stellar-scaled.

Critically, whilst the average $v_{\rm{c}}/\sigma_{0}$ as a function of radius remains approximately constant for the stellar-scaled curve, the self-scaled curve is strongly biased toward low $v_{\rm{c}}/\sigma_{0}$ systems at large radii. This is potentially problematic as it means that different types of galaxies dictate the shape of the rotation curve at different radii. The self-scaled median rotation curve therefore cannot be deemed representative of the average for the {\it whole} sample. 

We have demonstrated that the self-scaled normalisation prescription leads to a bias in the types of galaxies contributing to the median average rotation curve at different radii. To understand whether this is of importance for our analysis, we must also understand the origins and effects of this bias. The lowest three panels of Figure~\ref{fig:biascheck} show that the shape and extent of the self-scaled curves change as a function of $v_{\rm{c}}/\sigma_{0}$. For galaxies with the lowest values of $v_{\rm{c}}/\sigma_{0}$ we are able to trace their rotation curves out to larger multiples of the scale radius ($R_{\rm{t}}$). At the same time, it is only these low $v_{\rm{c}}/\sigma_{0}$ systems that exhibit an obvious decline in the outer parts of their (scaled) rotation curves. The self-scaled rotation curves of galaxies with higher $v_{\rm{c}}/\sigma_{0}$ do not extend out far enough to tell whether they remain flat or turn over too, being entirely compressed to within $\pm1R_{\rm{t}}$. Thus the bias in the self-scaled median curve {\it does} impact on its shape at large scale radii. Conversely, the stellar-scaled curves remain comparatively constant in both shape (remaining flat or continuing to rise) and radial extent with changing $v_{\rm{c}}/\sigma_{0}$. 

The {\it origin} of the bias in the self-scaled curve is straightforward to explain: first, we expect galaxies with smaller $R_{\rm{t}}$ to be disproportionately represented in the outer parts of the curve; if $R_{\rm{t}}$ is small then the data are more likely to trace the rotation curve out to larger multiples of this smaller value. Second, in Appendix~\ref{sec:stackbias} we show that KROSS galaxies with the lowest $v_{\rm{c}}/\sigma_{0}$ values also have much smaller sizes than the median size for the sample. This effect is seen in both $R_{\rm{t}}$ and $R_{\rm{d}}$, but is strongest in $R_{\rm{t}}$. Thus the selection for low $v_{\rm{c}}/\sigma_{0}$ systems at larger radii in the self-scaled curve is actually a selection for galaxies with small values of $R_{\rm{t}}$. Furthermore in Appendix~\ref{sec:stackbias} we show that, for a sub-set of 102 KROSS galaxies for which \citet{vanderWel:2012} S\'ersic index measurements are available, those galaxies that are dispersion-dominated ($v_{\rm{c}}/\sigma_{0} < 1$) have a higher median average S\'ersic index than the median for rotation-dominated ($v_{\rm{c}}/\sigma_{0} > 1$) systems, or for the total sub-set. 

Having established that the outer shape of the average self-scaled rotation curve is entirely dictated by small (low $R_{\rm{t}}$), dispersion-dominated (low $v_{\rm{c}}/\sigma_{0}$) galaxies with higher than average S\'ersic indices, one might assume then that the self-scaled prescription preferentially selects for systems with a more prominent bulge component to contribute toward the average curve at larger multiples of the scale radius. These systems would be more likely to exhibit declines in their individual rotation curves since they do not have high levels of circular motion and therefore violate the assumption that the rotation velocity is an effective probe of the dynamical mass. Furthermore, in these cases $R_{\rm{t}}$ is anyway small, so using it as a scaling factor acts to effectively ``zoom in'' on a small {\it physical} region of the galaxy, despite this region extending out to larger multiples of $R_{\rm{t}}$. Assuming these systems are indeed more bulge dominated, the radial scaling means only a region of the curve that is, by definition, keplarian in its shape is considered. 

This ``zoom'' interpretation is further supported by the fact that the stellar-scaled curves do not exhibit the same behavior as the self-scaled curves at low $v_{\rm{c}}/\sigma_{0}$, suggesting that $R_{\rm{t}}$ and $R_{\rm{d}}$ are measuring very different physical scales within a galaxy in the low $v_{\rm{c}}/\sigma_{0}$ regime. Indeed, the median $R_{\rm{d}}/R_{\rm{t}}$ for KROSS galaxies in the stack with $v_{\rm{c}}/\sigma_{0} < 1$ is $45\pm26$ percent larger than the median for the total KROSS stack sample. This suggests $R_{\rm{d}}$, measured from broadband imaging rather than the rotation curve itself, is likely less sensitive than $R_{\rm{t}}$ to the presence of a bulge-like component and better reflects the overall spatial extent of the stellar light (and thus mass) of the entire galaxy (rather than a single component).

Now considering instead those galaxies with higher $v_{\rm{c}}/\sigma_{0}$ (i.e.\ the majority of the galaxies) one might expect, given the evidence discussed, that these systems have a less prominent bulge component or none entirely and thus a more smoothly rising or flat rotation curve. It is therefore unsurprising that the best fit $R_{\rm{t}}$ for these systems is indeed preferentially found towards the maximum radial extent of the data. The result is a compression of the majority of the scaled curves in the sample to within $\pm1R_{\rm{t}}$. Thus our earlier conjecture on the risks of scaling the rotation curves by $R_{\rm{t}}$ is proved correct. 

In summary then the self-scaled scaling prescription effectively compresses the majority of the scaled rotation curves in the sample to within $\pm1R_{\rm{t}}$, leaving only a minority of low $v_{\rm{c}}/\sigma_{0}$, small $R_{\rm{t}}$ galaxies (with higher than average S\'ersic indices) to dictate the outer shape of the final average scaled curve. We thus conclude that the stellar-scaled stacked rotation curves provide a fairer representation of the typical rotation curve shape than the self-scaled stacked curves for the galaxy samples. We thus proceed to adopt the stellar-scaled scaling prescription for the remainder of our analysis and do not discuss the self-scaled curves any further. 

\subsection{Stacking Nebular Emission}
\label{subsec:specs}

The radial extent of the median stack of the individual rotation curves is limited by the spatial extent of detected H$\alpha$ emission in each individual galaxy; for galaxies to contribute to the median stack at a given radius, they must individually have detectable levels of H$\alpha$ emission at that radius. Using this method we were able to trace the KROSS average rotation curve out to $\approx6R_{\rm{d}}^{\prime}$. However, for our other redshift bins we could only measure the average curve out to $\approx4R_{\rm{d}}^{\prime}$ (equivalent to $\approx$9 kpc, or a radius containing $\approx90$ percent of the total stellar mass for a pure exponential disk). Furthermore, even for the more extended KROSS curve, with increasing scaled radius the average normalised velocity is measured from an ever decreasing number of galaxies in the sub-sample. This means that the shape of the rotation curve is, by definition, strongly biased toward those systems with the largest H$\alpha$ flux at large multiples of the scale radius.

To overcome these limitations we instead stack the galaxies' nebular emission itself in the form of position-velocity diagrams normalised in radius, velocity and flux. Stacking the position-velocity diagrams allows for faint nebular emission, undetected in a given data cube spaxel, to contribute to the final average rotation curve. This allows us to construct average rotation curves that extend beyond $\approx4R_{\rm{d}}^{\prime}$ at each redshift, and that are more representative of the typical rotation curve shape for the sample at all radii.

In this sub-section we describe the methods used to construct these position-velocity diagrams for each of the integral field spectroscopy data cubes in our sample. We detail how we stack the diagrams and subsequently extract an average rotation curve from each stack. Since the ultimate goal of this work is to measure the average dark matter fraction within as close as possible to the total extent of the stellar mass for each of our galaxy samples, we aim to measure the average rotation curve out to at least $\approx6R_{\rm{d}}^{\prime}$ ($\approx$13 kpc) in each of our redshift bins. This radius should encompass $\approx98$ percent of the total stellar mass (assuming a pure exponential disk profile), allowing a measure of the total dark matter fraction within the spatial extent of the starlight. This radius is also similar to the maximum radii of existing H$\alpha$ rotation curves measured for galaxies in the local Universe \citep[e.g.][]{Catinella:2006,Catinella:2007}. 

\subsubsection{Average Position-Velocity Diagrams}
\label{subsec:extractstack}

To construct a position-velocity diagram for each galaxy we first extract a series of spectra along each galaxy's major kinematic axis. For each galaxy we identify the major kinematic axis (see \S~\ref{subsec:linemaps}) and extract spectra from the cube by summing the flux from circular bins placed along this axis, spaced in multiples of $R_{\rm{d}}$, each time added in quadrature with the $\sigma$ width of the PSF i.e. the same ($R_{\rm{d}}^{\prime}$, $v_{3R_{\rm{d}}^{\prime}}$) stellar-scaling described in \S~\ref{subsec:stackRC}. Each bin has a width equal to the FWHM of the PSF associated with the cube. For each spectrum we convert the wavelength axis values, $\lambda_{\rm{i}}$ in to line-of-sight velocities as $v_{\rm{i}} = c(\lambda_{\rm{i}}-\lambda_{\rm{H}\alpha})/\lambda_{\rm{H}\alpha}$, where $\lambda_{\rm{H}\alpha}$ is the observed central wavelength of the H$\alpha$ emission within the central bin as determined from a triple-Gaussian fit to the H$\alpha$ and [N{\sc ii}] emission. This should correspond to the rest-frame wavelength of H$\alpha$ multiplied by $1+z$, where $z$ is the redshift of the galaxy. For the MUSE cubes, we perform a double-Gaussian fit to the [O{\sc ii}] doublet in the central spectrum, similarly converting the wavelength axis values to line-of-sight velocities.

For each galaxy we use a linear interpolation to re-sample the spectra on to a common, uniform grid of normalised radius and velocity to produce a position-velocity diagram. The radius and velocity scalings are in units of $R_{\rm{d}}^{\prime}$ and $v_{3R_{\rm{d}}^{\prime}}$, respectively. The pixel size of the grid (steps of 0.25 in $R_{\rm{d}}^{\prime}$, and 0.15 in $v_{3R_{\rm{d}}^{\prime}}$) is chosen as a compromise between maximising the signal-to-noise of the nebular emission in each pixel and the ability to accurately centre each diagram for stacking. To produce an average position-velocity diagram we first normalise each individual diagram by the average of its peak flux at $r  = \pm3R_{\rm{d}}^{\prime}$. We self-normalise the flux of the diagrams in this manner to avoid preferentially biasing the stack towards the brightest galaxies (that will also be the most massive, on average)\footnote{Our tests show that normalising the diagrams in this manner only changes the outer slope of the final rotation curve extracted from the stacked diagrams by $5\pm4$ percent in comparison to if no normalisation is applied.}. Furthermore, we choose to normalise by the flux at $3R_{\rm{d}}^{\prime}$, rather than the central ($r=0$) flux, to avoid preferentially biasing the stack toward galaxies that are more centrally concentrated. Finally, we combine the total set of normalised diagrams via a median average. To avoid giving undue weight to noise in the diagrams, we exclude from our average all those pixels in each individual diagram with a normalised flux value less than 1 percent.

To increase the signal-to-noise ratio of the nebular emission in the final stack, we also construct ``wrapped'' stacked position-velocity diagrams; we ``wrap'' or fold each individual position-velocity diagram about its origin in both radius and velocity space, taking the median average of the diagrams either side of the fold. We then median combine the total set of these wrapped diagrams in the same manner as described above. The final stacked diagram is median-filtered with a kernel of five pixels \citep[in line with the methods of][]{Lang:2017}. 

\subsubsection{Extracting Rotation Curves}
\label{subsec:extractrot}

To extract a rotation curve from each median stacked position-velocity diagram, we require a measure of the peak velocity at each radius. We therefore perform a fit to the flux in the diagram along each pixel column (i.e. to each spectrum at each radius increment). 

For the stacked H$\alpha$ emission, we parameterise the shape of the stacked flux by the sum of two Gaussian profiles: a broad, low-amplitude Gaussian that describes the stack of any continuum emission present in the individual spectra (plus any dispersed [N{\sc ii}] emission); and a second, narrow and higher amplitude Gaussian to describe the stacked H$\alpha$ emission.

For the stacked [O{\sc ii}] emission, the relative amplitude of the two doublet lines is much closer to unity than that of H$\alpha$ and [N{\sc ii}] emission. Stacking the [O{\sc ii}] emission from different galaxies therefore produces a skewed Gaussian profile, superimposed on the broader, Gaussian-profile continuum emission. We therefore avoid any interpretation of the MUSE stacked position-velocity diagrams until after they have been ``wrapped'', at which point the skewed Gaussian shape instead becomes symmetrical and can be described in the same way as the H$\alpha$. 

To measure the velocity and its uncertainty in the extracted rotation curve, we bootstrap the median position-velocity diagram, repeatedly selecting an equally sized, random sample of the individual galaxy position-velocity diagrams before median combining them and extracting a rotation curve. We repeat this process 100 times, taking the median and median absolute deviation (with respect to the median itself) of each of the 100 extracted curves at each radius as respectively the velocity and its uncertainty.

\subsubsection{Cross-verification of Methods and Further Checks}
\label{subsubsec:crossverify}

To check the validity of the rotation curves extracted from our stacked position-velocity diagrams in comparison to the median stack of the individual rotation curves, in Appendix~\ref{subsec:pvversusRS} we verify that the former agrees with the latter for the KROSS sample, showing the two curves agree within uncertainties. We can therefore be confident in the accuracy of rotation curves extracted from the stacked position-velocity diagrams, and that the shape of the average rotation curve does not significantly differ as a function of the method of its construction.

We also note here that, as described in \S~\ref{subsubsec:stellarsizes}, our calculation of the stellar disk-scale radius is based on the assumption that $R_{\rm{d}}=0.59R_{\rm{h}}$. Of course this is only strictly true for a pure exponential disk, i.e. with a sersic index $n=1$. However, the median sersic index is $1.1\pm0.5$ for those 356 galaxies in our samples with a \citet{vanderWel:2012} sersic index measurement so this is a reasonable assumption for this work. Nevertheless, we verified that our stacked position-velocity diagrams are not biased by the inclusion of galaxies with a sersic index $n\neq1$ by dividing those 356 galaxies into four bins of sersic index and producing stacked position-velocity diagrams and rotation curves for the galaxies in each bin. We find no trend, within 1$\sigma$ uncertainties, between the outer slopes (the ratio of the velocity at $6R_{\rm{d}}^{\prime}$ to that at $3R_{\rm{d}}^{\prime}$; see \S~\ref{subsubsec:quantRCshape}) of the extracted rotation curve in the four bins, nor in comparison to the outer slope of the rotation curve derived from the total stacked position-velocity diagram of all 356 galaxies. 

Similarly, in \S~\ref{subsubsec:stellarsizes} we noted that a subset of (KGES) galaxies in our sample were also analysed by \citet{vanderWel:2012}, and that on average our measurements of $R_{\rm{h}}$ agreed well with theirs, but with a (small) scatter of $\pm0\farcs1$. Assuming we may expect a similar level of scatter in $R_{\rm{h}}$ between our samples as a results of slight differences in methodology, we used the KROSS sample to quantify how this expected variation may affect our final rotation curve. For a systematic change of $\pm0\farcs1$ in $R_{\rm{h}}$ we found no significant difference in the outer slope of the rotation curve extracted from the stacked position-velocity diagram of KROSS galaxies.  

\subsubsection{Modelling Rotation Curves}
\label{subsec:modelstack}

To trace the shape of each rotation curve extracted from the stacked position-velocity diagrams we find the best fit model to each. To avoid biasing our conclusions as a result of the choice of model we fit several different functional forms, and use the {\it Akaike information criterion} \citep[AIC; e.g.][]{Akaike:1998} to select the most appropriate, accounting for the goodness of fit (i.e. the $\chi^{2}$ value), the number of data points and the number of free parameters in the model. We determine the best choice model as the one with the lowest AIC number. 

To each rotation curve we fit three commonly employed models: the \citet{Courteau:1997aa} arctangent disk model for galaxy rotation curves, an exponential disk model, and lastly a model comprising the sum of a normalised exponential disk and a pseudo-isothermal dark matter halo. The analytical forms of the three models are described in Appendix~\ref{sec:RCfittingfunctions}.

Since in all cases we fit to normalised rotation curves, we do not physically interpret the best fit parameters of any of the three models but rather use them only as a convenient means to recover the intrinsic shape of the curves after accounting for the noise, size of the data set and complexity of the model.

\begin{figure*}
\centering
\begin{minipage}[]{1.\textwidth}
\centering
\includegraphics[width=1.01\textwidth,trim= 20 0 0 0,clip=True]{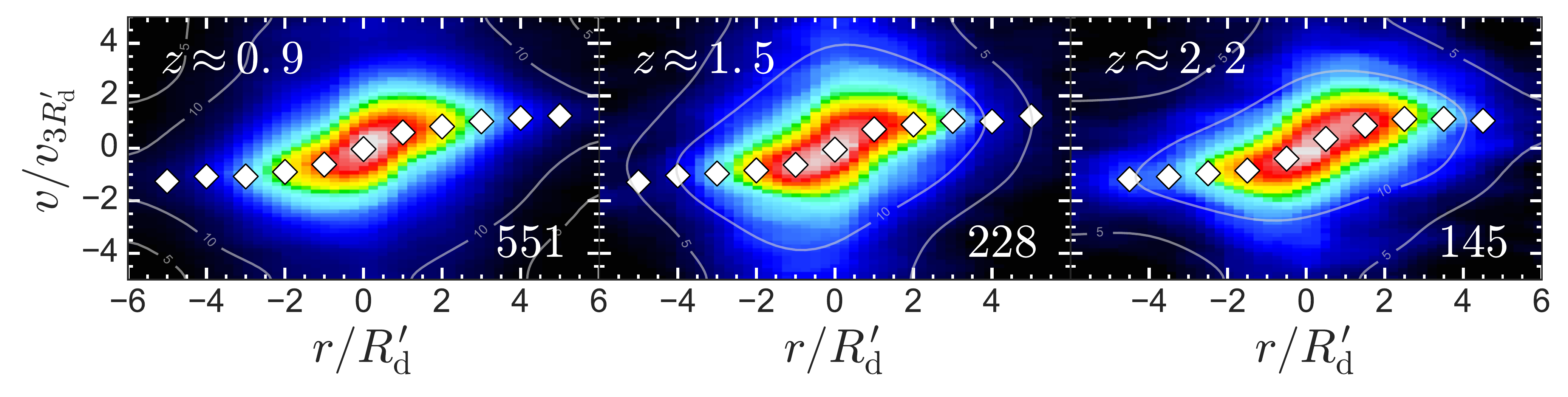}
\end{minipage}
\caption{%
The median normalised position-velocity diagram for the three KMOS samples, constructed via the methods detailed in \S~\ref{subsec:specs}. The linear colour scale represents the normalised flux intensity from black at the lowest flux values to white at the highest flux values. We overlay grey contours corresponding to a signal-to-noise ratio of 5 and 10. For each position-velocity diagram we extract the normalised velocity at each radius via a fit to the spectrum from the corresponding radial bin in the stacked position-velocity diagram (see \S~\ref{subsec:extractstack}). This curve is represented in each case by white diamond points. Each of the rotation curves either remains approximately flat or continues to rise with increasing radius out to $6R_{\rm{d}}^{\prime}$.%
     }%
\label{fig:total_stacks}
\end{figure*}

\begin{figure*}
\begin{minipage}[]{1.\textwidth}
\includegraphics[width=1.01\textwidth,trim= 25 10 5 0,clip=True]{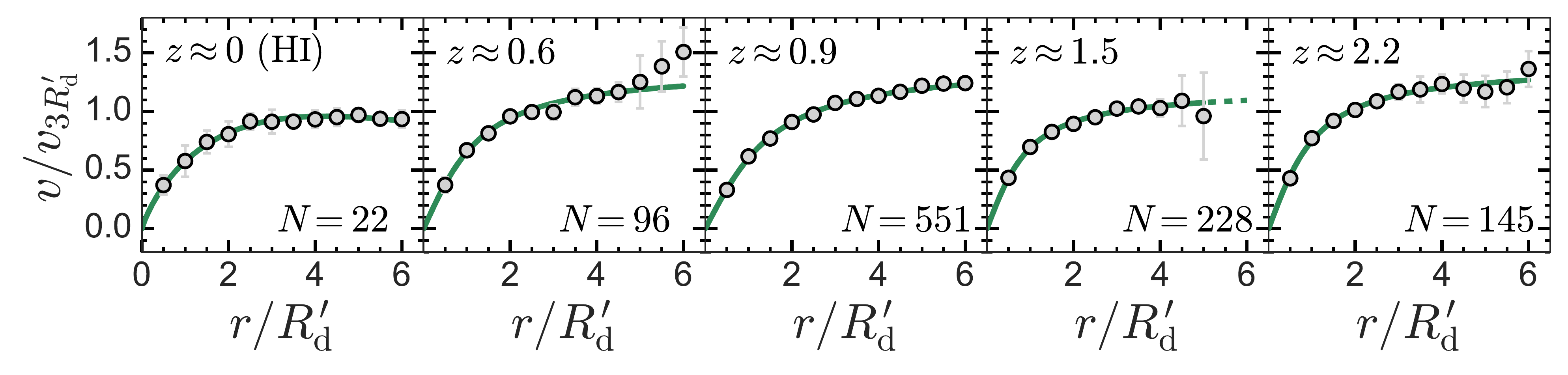}
\end{minipage}
\caption{%
Normalised, wrapped median galaxy rotation curves as a function of redshift for our galaxy samples. For the four right-most panels, the velocity at each radius is extracted via a fit to the spectrum from the corresponding radial bin in the stacked position-velocity diagram. For the $z\approx0$ panel, the rotation curve is constructed via a median stack of the wrapped, H{\sc i}-derived rotation curves for each $z\approx0$ galaxy. The best fit model rotation curve to the data in each case is presented via a green solid line (or a green dashed line where the model is extrapolated beyond the data). Each of the rotation curves either remains flat or continues to rise out $6R_{\rm{d}}^{\prime}$. %
     }%
\label{fig:total_wrapped_curves}
\end{figure*}

\subsubsection{Quantifying the Shape of Rotation Curves}
\label{subsubsec:quantRCshape}

To quantify, and draw comparisons between, the shapes of the average rotation curves constructed in our analysis, we devise a simple one parameter measure of the extent to which each curve declines at large radii. This is simply the ratio of the velocity measured at $6R_{\rm{d}}^{\prime}$ and that measured at $3R_{\rm{d}}^{\prime}$, such that the ``turnover'', $t = v_{6R_{\rm{d}}^{\prime}}/v_{3R_{\rm{d}}^{\prime}}$. In this respect, if a galaxy rotation curve remains flat $t=1$, if it is rising $t>1$, and if it is falling $t<1$. We choose $6R_{\rm{d}}^{\prime}$ ($\approx$13 kpc, on average) and $3R_{\rm{d}}^{\prime}$ ($\approx$6.5 kpc, on average) as respectively the typical maximum radius that we are able to probe in our average rotation curves, and the radius that should be slightly beyond the rotation curve maximum.

We expect this outer slope to be linked to the average dark matter fraction of the individual galaxies that contribute to the average curve. This is discussed further in \S~\ref{subsec:estimate_fDM}, where we formally link the two via a comparison with a toy model. However, for the bulk of our analysis we proceed to compare the shapes of average galaxy rotation curves in terms of $t$.

\section{Results and Discussion}
\label{sec:results}

In this section we present median, stellar-scaled position-velocity diagrams and the corresponding normalised rotation curves derived from stacked nebular emission from samples of star-forming galaxies at $z\approx0.6$ to $z\approx2.2$. As a baseline for our results, we also include rotation curves constructed from the median average H{\sc i}-derived rotation curves from THINGS galaxies at $z\approx0$. We explore the extent to which the outer slopes of the galaxy rotation curves correlate with intrinsic galaxy properties and compare this to the trends observed for model galaxies in the EAGLE simulation.\footnote{We note that, whilst our results and discussion concern only stellar-scaled rotation curves, in Appendix~\ref{subsec:recoveraturn} we show that we are able to recover a rotation curve that significantly declines at large scale radii if we adopt the self-scaled normalisation prescription \citep[in agreement with Figure~5,][]{Lang:2017}. However, we do not interpret these results physically due to the biases inherent in the self-scaled curves, as discussed in \S~\ref{subsec:stackRC}.} From our results we also calculate an estimate of the average dark matter fraction of star-forming galaxies as a function of redshift since $z\approx2.2$. 

\subsection{Total Stacks}
\label{subsec:totalstacks}

Figure~\ref{fig:total_stacks} illustrates the distribution of stacked flux in the normalised position-velocity plane for our full KMOS samples, with signal-to-noise contours overlaid. The noise in each pixel of the stack is calculated as $\sigma_{\rm{noise}}=1/\rm{N}_{\rm{d}} \times \sqrt{\sum_{i=1}^{\rm{N}_{\rm{d}}} \sigma_{i}^{2}}$, where $\sigma_{i}$ is the uncorrelated noise in each of the $\rm{N}_{\rm{d}}$ individual position-velocity diagrams that effectively contribute to the pixel, measured in each case in a region of the diagram far from any potential emission. There is no strong decline in rotation velocity apparent at large radii in the position-velocity diagram for any of the three samples. Before exploring the shapes of the rotation curves in more detail we first boost the signal in the final stacks by stacking instead the {\it wrapped} position-velocity diagrams for the galaxies in each sample, as described in \S~\ref{subsec:extractstack}. The rotation curves extracted from the wrapped stacks are shown in Figure~\ref{fig:total_wrapped_curves}. Here we also include the curves constructed from the stacked [O{\sc ii}] emission from the total MUSE sample, and from the median average of the H{\sc i}-derived THINGS rotation curves. Each of the wrapped rotation curves, measured within $\lesssim 6R_{\rm{d}}^{\prime}$, either remains flat or rises slightly with increasing radius ($t=1.00\pm0.02$, $t=1.14\pm0.03$, $t=1.16\pm0.01$, $t=1.10\pm0.03$, and $1.11\pm0.03$ for respectively the $z\approx0$, $z\approx0.6$, $z\approx0.9$, $z\approx1.5$, and $z\approx2.2$ average curve).

\subsection{Binned Stacks}
\label{subsec:turnorigins}

\begin{figure*}
\centering
\begin{minipage}[]{1\textwidth}
\centering
\includegraphics[width=1.01\textwidth,trim= 40 0 25 10,clip=True]{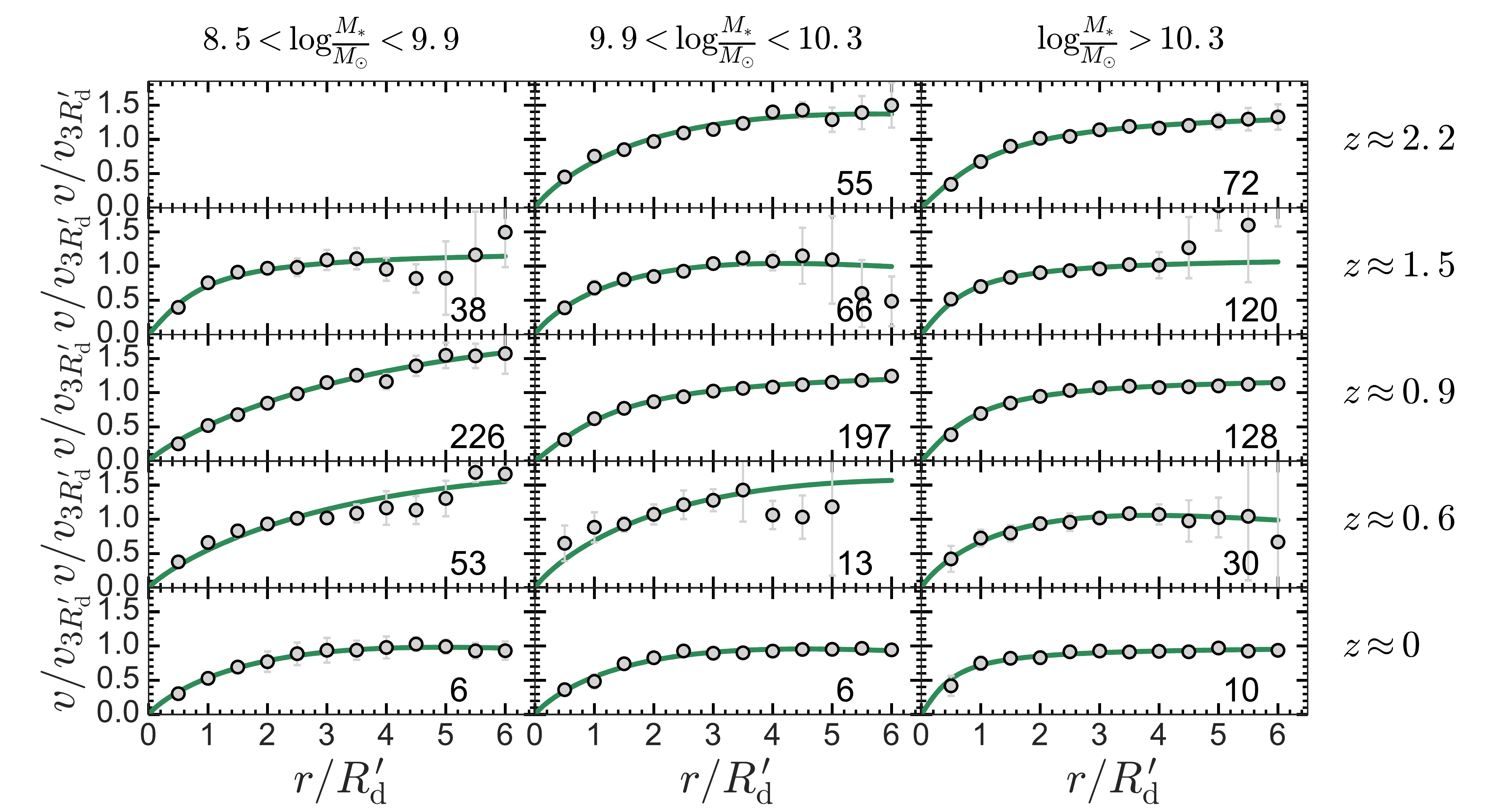}
\end{minipage}
\caption{%
Average wrapped, normalised rotation curves for samples of galaxies from our analysis separated in to bins of redshift and mass. The rotation curves were constructed in the same manner as for Figure~\ref{fig:total_wrapped_curves}. The green solid line in each panel represents the best fit model curve. The shape of each rotation curve is consistent, within uncertainties, with remaining flat or continuing to rise out to large radii.%
     }%
\label{fig:gridplot_curves}
\end{figure*}

\begin{figure*}
\centering
\begin{minipage}[]{1\textwidth}
\centering
\includegraphics[width=.85\textwidth,trim= 0 0 0 0,clip=True]{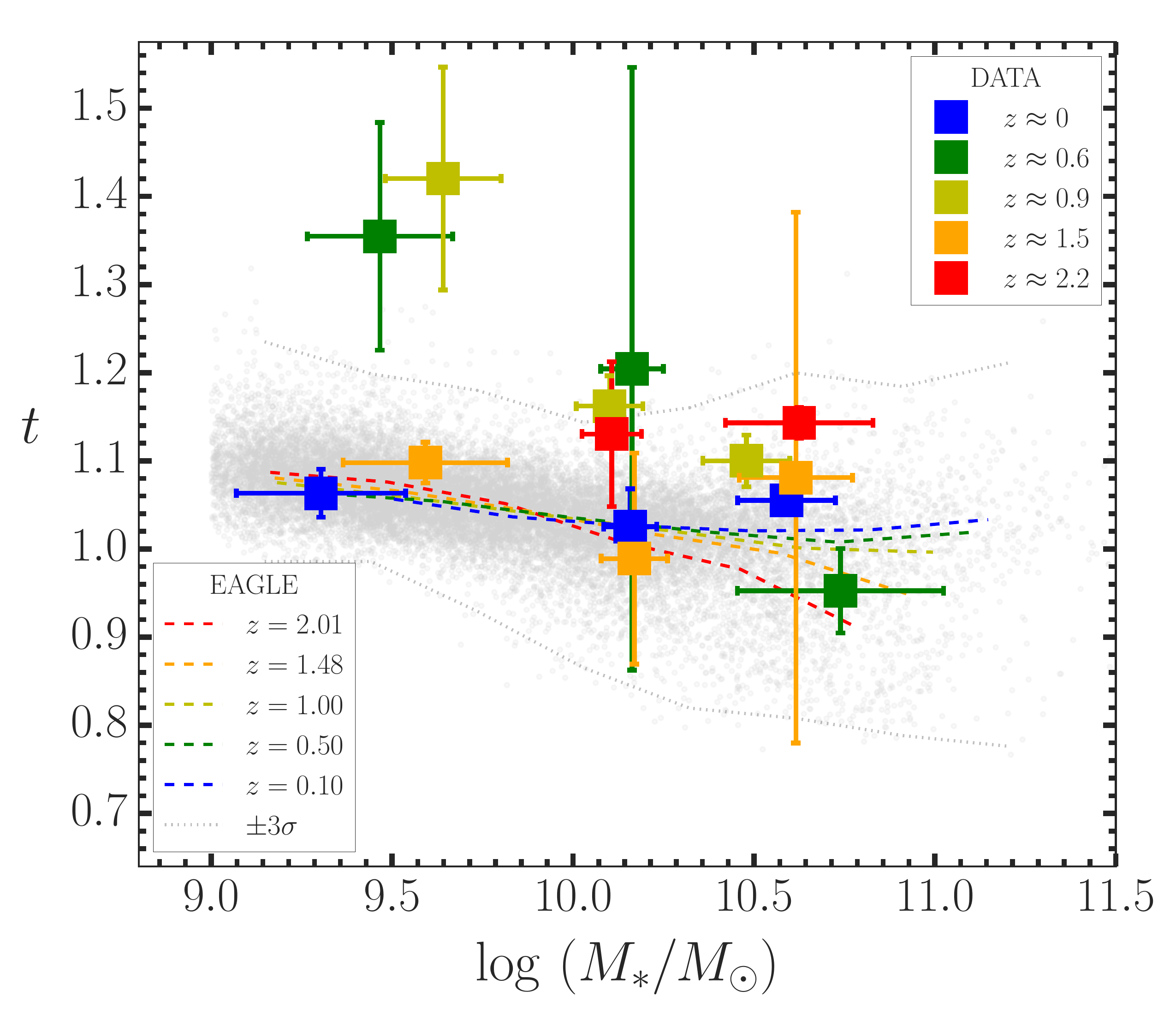}
\end{minipage}
\caption{%
Galaxy rotation curve turnover, $t = v_{6R_{\text{d}}^{\prime}}/v_{3R_{\text{d}}^{\prime}}$, as a function of stellar mass. Measurements for model EAGLE galaxies at $z=$0.1, 0.5, 1.0, 1.48, and 2.01 are displayed in grey. The three-sigma envelope for the EAGLE galaxies is indicated via dotted grey lines. The running median for each redshift in EAGLE is shown as a coloured dashed line. Overlaid as squares (colour-coded according to the median redshift) are the measurements for each of the stacked rotation curves shown in Figure~\ref{fig:gridplot_curves}, where $t$ for each stack (i.e. for each panel in Figure~\ref{fig:gridplot_curves}) is measured from the best fit model to the rotation curve. There is only a weak trend in turnover with stellar mass in either the observed data or for model galaxies in EAGLE, with the two in general agreement within uncertainties. The observed rotation curves are, in every case, consistent with being flat or continuing to rise between $3R_{\rm{d}}^{\prime}$ and $6R_{\rm{d}}^{\prime}$. Some EAGLE galaxies do display falling rotation curves, down to $t\approx0.8$. These are in a minority in each of the EAGLE redshift slices, but they do represent a larger fraction with increasing lookback time; less than 1 percent of model EAGLE galaxies at $z\leq0.5$ have a $t<0.9$, whereas $2$--$4$ percent do at $z\geq1$. 
     }%
\label{fig:massturn_eagle}
\end{figure*}

\begin{figure*}
\centering
\begin{minipage}[]{1\textwidth}
\centering
\includegraphics[width=.85\textwidth,trim= 0 0 0 0,clip=True]{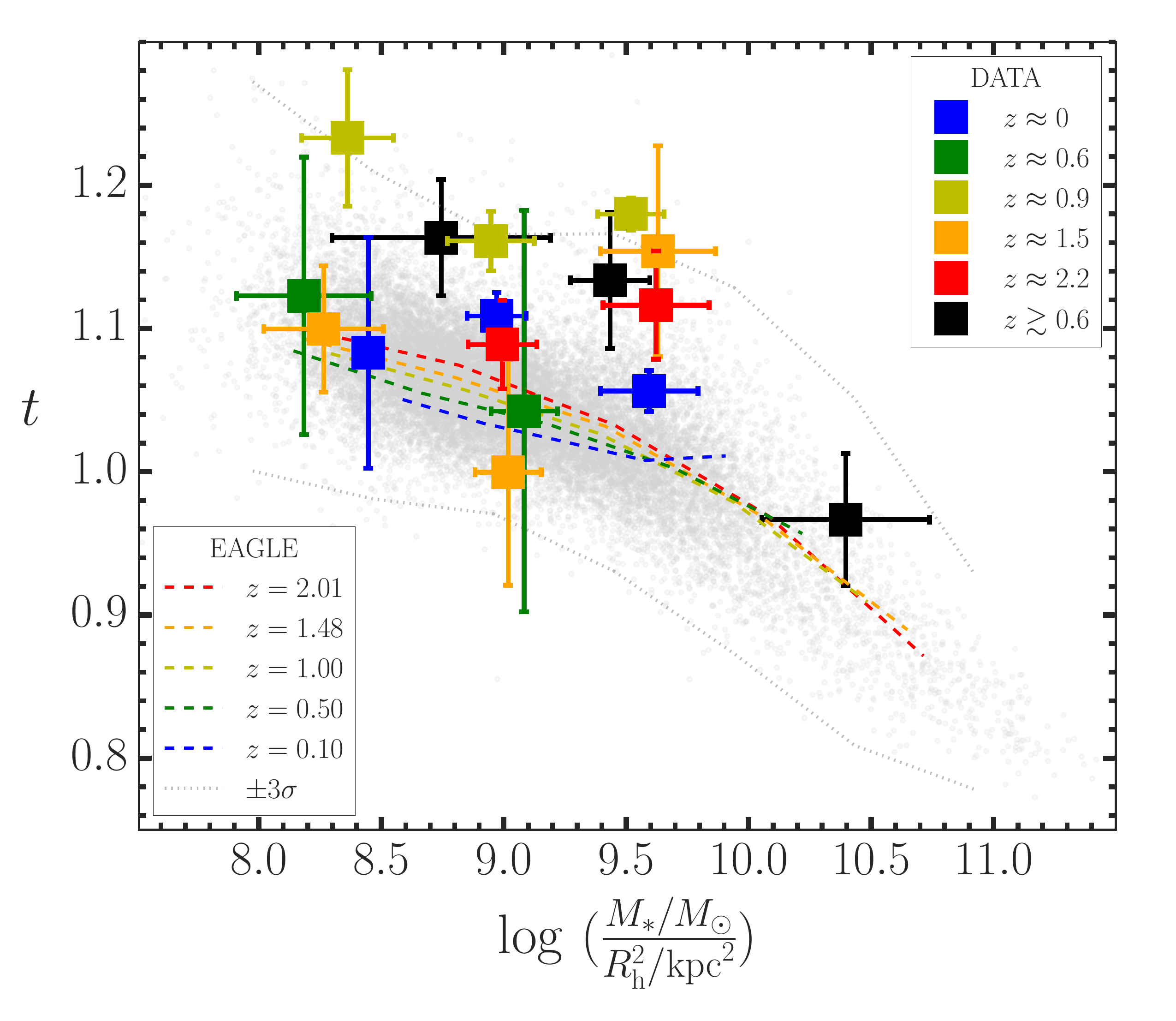}
\end{minipage}
\caption{%
Galaxy rotation curve turnover, $t = v_{6R_{\text{d}}^{\prime}}/v_{3R_{\text{d}}^{\prime}}$, as a function of stellar mass surface density. Measurements for model EAGLE galaxies at $z \leq 2.01$ are displayed in grey. The median and three-sigma envelope for the EAGLE galaxies are indicated via dotted grey lines. Those EAGLE galaxies with the highest stellar mass surface density have the strongest turnovers in their rotation curves. Further, those EAGLE galaxies with the highest surface mass densities (and thus most strongly falling rotation curves) are most frequent at $z \approx 2$. Overlaid as squares (colour-coded according to the median redshift) are the measurements for each of the stacked rotation curves shown in Figure~\ref{fig:gridplot_curves_dens}, where $t$ for each stack (i.e. for each panel in Figure~\ref{fig:gridplot_curves_dens}) is measured from the best fit model to the rotation curve. Measurements of $t$ measured for the total combined ($z\gtrsim0.6$) KMOS sample split in to three density bins are also shown. The observed trend in turnover with stellar mass surface density is consistent with the trend from EAGLE, within uncertainties. There is no evidence for any significant deviation from the expectations of $\Lambda$CDM.%
    }%
\label{fig:gridplot_eagle_obs}
\end{figure*}

\begin{figure*}
\centering
\begin{minipage}[]{1\textwidth}
\centering
\includegraphics[width=1.\textwidth,trim= 0 0 0 0,clip=True]{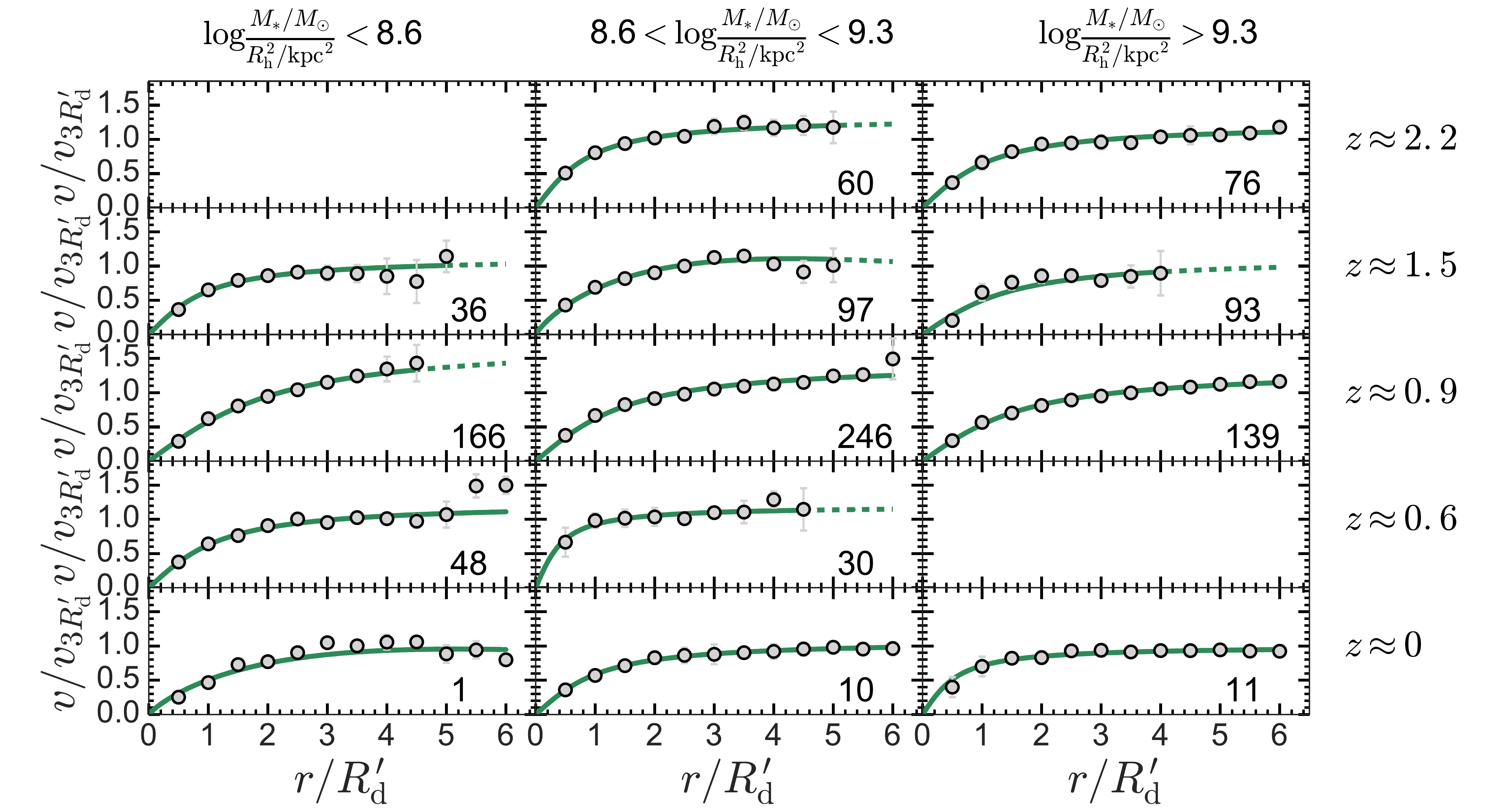}
\end{minipage}
\caption{%
Average wrapped, normalised rotation curves for samples of galaxies from our analysis as a function of redshift (vertical) and central stellar mass surface density (horizontal). The rotation curves were constructed in the same manner as for Figure~\ref{fig:total_wrapped_curves}. The green solid line in each panel represents the best fit model curve. The line is dashed when the model is extrapolated beyond the data. For all bins, the rotation curves either remain approximately flat or continue to rise out to the maximum radius probed in each case.%
     }%
\label{fig:gridplot_curves_dens}
\end{figure*}

In \S~\ref{subsec:totalstacks}, we showed that the average rotation curve in each of our redshift bins either remains flat or continues to rise out to $\approx6R_{\rm{d}}^{\prime}$. Each of these stacks contains a large number of galaxies,  representing the average of many hundreds of systems. In this sub-section, we investigate whether smaller sub-samples of galaxies exist with average rotation curves that decline in their outskirts, or whether flat or rising rotation curves are ubiquitous across our star-forming galaxy samples across each redshift. We also compare our results to those for model galaxies from the EAGLE simulation.

\subsubsection{EAGLE}
\label{subsubsec:eaglerc}

To inform our interpretation of the results for our observed galaxy samples, we compare them to model galaxies from the EAGLE hydrodynamical cosmological simulation. We select sub-samples of star-forming ($>1M_{\odot}$ yr$^{-1}$ for a reasonable comparison to the data) model EAGLE galaxies in bins of increasing redshift ($z=$ 0.10, 0.50, 1.00, 1.48, and 2.05) chosen to cover a similar range in redshift to our observed galaxy samples, and with stellar masses $\log M_{*}/M_{\odot} \geq 9$.

We compute the rotation curves of the model EAGLE galaxies from the
simulation data using the method of \cite{Schaller:2015a}, briefly summarised here for completeness. Haloes are identified in the simulation using a friends-of-friends algorithm on the dark-matter structures. Self-bound sub-structures are then extracted using the \textsc{Subfind} algorithm \citep{Springel:2001,Dolag:2009}. The stellar
component of these bound structures corresponds to the individual galaxies. We compute their properties (e.g. stellar masses and star-formation rates) in spherical apertures of $30$~(physical) kpc, a
size chosen to match the Petrosian aperture used in SDSS photometry data \citep{Schaye:2015}. These quantities have been made publicly available in the form of a database \citep{McAlpine:2015}.

For each model galaxy we construct concentric spherical shells around the centre of potential of each sub-structure and obtain the mass enclosed in each shell $M(<r)$. This allows us to construct both a density profile and a circular velocity curve using $V_{\rm circ}(r) =\sqrt{GM(<r)/r}$. These rotation curves have been shown to be an excellent match to low-redshift observational data \citep[e.g.][]{Schaller:2015a,Ludlow:2017} over a wide range of galaxy stellar masses. We stress here that the circular velocities are derived from the potential, rather than representing directly the velocities of the matter. 

The simulations have been shown to produce converged stellar masses for all galaxies above $10^8\, M_{\odot}$ and converged sizes and star formation rates for objects with a mass above $10^9\, M_{\odot}$
\citep{Schaye:2015}. The rotation curves are converged at better than the $10$ percent level at radii larger than $2$--$3$\, kpc. Once stacked, this limit shrinks and the rotation curves have been show to be well converged at all radii larger than $1$\, kpc \citep{Schaller:2015a}. The simulations are hence well matched to the observational data used in this work. 

\subsubsection{Rotation Curve Shape Versus Stellar Mass}
\label{subsubsec:rcvsmass}

First we examine the rotation curves derived from our star-forming galaxy samples split in to bins of stellar mass and redshift. In Appendix~\ref{subsec:morebinpvdiags} we show the median stacked position-velocity diagrams for star-forming galaxies from our samples separated into bins of stellar mass and redshift to demonstrate that the position-velocity diagrams are still well-behaved after having split our sample in to smaller sub-samples. To boost the signal-to-noise, we again construct stacks of the wrapped position-velocity diagrams. These are shown in Figure~\ref{fig:gridplot_curves}, where we also include the wrapped curves for the MUSE and THINGS samples.

From Figure~\ref{fig:gridplot_curves} we see that all of the rotation curves remain flat, or continue to rise, out to $\approx6R_{\rm{d}}^{\prime}$, suggesting that star-forming galaxies with stellar masses $\log M_{*}/M_{\odot} > 8.5$ at all redshifts contain substantial amounts of dark matter within this radius. The average rotation curves in the central mass bin ($9.9 < \log M_{*}/M_{\odot} < 10.3$) for $z\approx1.5$ and $z\approx0.6$ do qualitatively exhibit declines in their outskirts but with large scatter. However, after calculating the AIC for the best fit to the data for each of the three different functional forms described in \S~\ref{subsec:modelstack}, we safely reject the possibility that these declines are real; the preferred best fit model remains approximately flat or continues to rise in each case. 

To quantify the rotation curve shape we calculate the rotation curve turnover, $t=v_{6R_{\rm{d}}^{\prime}}/v_{3R_{\rm{d}}^{\prime}}$ (see \S~\ref{subsubsec:quantRCshape}) in each of the stellar mass and redshift bins and plot these values in Figure~\ref{fig:massturn_eagle}. In this figure we identify a weak overall trend between stellar mass and $t$ for our samples, albeit with significant scatter. Importantly, there is no evidence to suggest a significant deviation from a flat or rising rotation curve within $6R_{\rm{d}}^{\prime}$ in any bin; the weighted average turnover for our sample with $z\approx0.6$--$2.2$ is $t=1.14\pm 0.02$. This suggests that the dark matter fraction of massive ($\log M_{*}/M_{\odot} \gtrsim 9$), star-forming galaxies at each redshift is similar, and does not vary strongly as a function of their stellar mass. 

Of course, one should consider whether we should {\it expect} any trend between rotation curve turnover and stellar mass (and redshift) in our sample, according to $\Lambda$CDM. To investigate this, we examine the trends for model galaxies in EAGLE. Unlike our observed samples, for the EAGLE galaxies we benefit from measurements of the rotation curves of {\it individual} systems out to large radii as well as having detail on their intrinsic properties. We therefore repeat the same experiment as for the observed galaxies in Figure~\ref{fig:massturn_eagle}, but for individual model EAGLE galaxies at each redshift (recalling that the absence of beam smearing effects in EAGLE means that $t = v_{6R_{\rm{d}}^{\prime}}/v_{3R_{\rm{d}}^{\prime}} \equiv v_{6R_{\rm{d}}}/v_{3R_{\rm{d}}}$ for the model EAGLE galaxies). In Figure~\ref{fig:massturn_eagle}, we include the results for star forming EAGLE galaxies in five different redshift slices, along with the running median for each slice. In general there is only a very weak trend between $t$ and stellar mass, although this trend is stronger for the highest redshift slice ($z\approx2$). The vast majority of model galaxies at all redshifts exhibit either flat or rising rotation curves ($t\gtrsim1$), with only a small minority exhibiting rotation curves that decline between $3R_{\rm{d}}$ and $6R_{\rm{d}}$ ($t<1$). This minority is comprised mostly of galaxies at higher redshift ($z\approx1.5$--$2$) in the EAGLE simulation. Additionally, it is only the most massive model galaxies at these epochs that exhibit such declining rotation curves.

\subsubsection{Rotation Curve Shape Versus Stellar Mass Surface Density}
\label{subsubsec:rcvsdens}

\begin{figure*}
\begin{minipage}[]{1\textwidth}
\centering
\includegraphics[width=0.95\textwidth,trim= 30 5 70 0,clip=True]{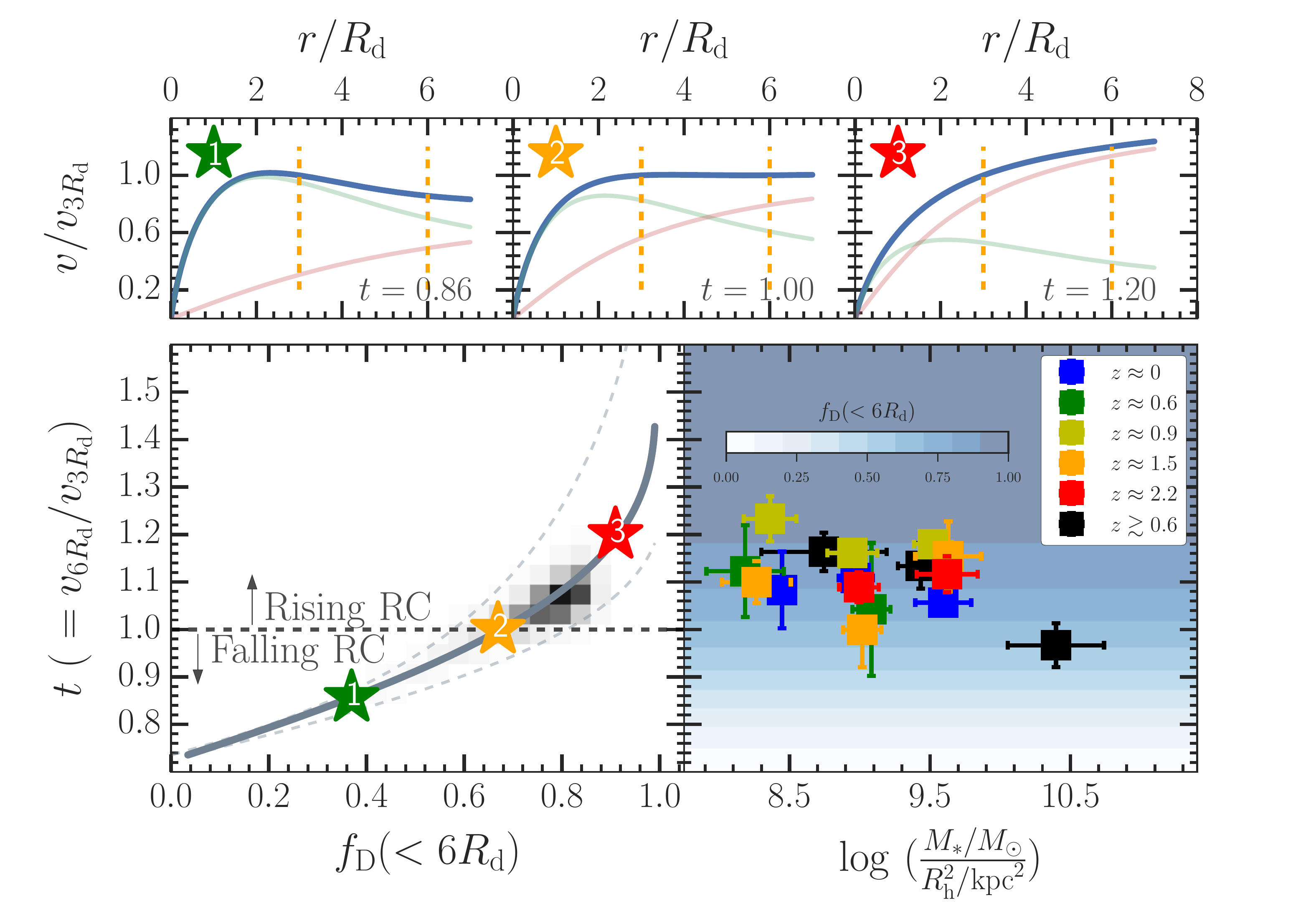}
\end{minipage}
\caption{%
{\bf Top:} Three example instances of the toy model, with numerical labels indicated by coloured stars. In each of three panels we include the total velocity as a solid blue line, and the contribution from respectively the baryonic disk and dark matter halo as a solid green line and a solid pink line. The dashed orange lines highlight the positions at which the rotation velocity is measured to calculate $t$ (i.e. $6R_{\rm{d}}$ and $3R_{\rm{d}}$). The panels show respectively a falling (top; $t < 1$), flat (middle; $t = 1$), and rising (bottom; $t > 1$) rotation curve. The position of each example on the $f_{\rm{D}}$-$t$ plane is indicated by their corresponding star in the Bottom Left plot. {\bf Bottom Left:} The dark mass fraction ($f_{\rm{D}}$) within $6R_{\rm{d}}$ as a function of galaxy rotation curve turnover, $t=v_{6R_{\rm{d}}}/v_{3R_{\rm{d}}}$ (where $v_{6R_{\rm{d}}}$ and $v_{3R_{\rm{d}}}$ are the rotation velocity at respectively $6R_{\rm{d}}$ and $3R_{\rm{d}}$) according to a simple exponential baryonic disk plus pseudo-isothermal dark matter halo toy model. The dashed grey lines indicate the bounding envelope for $10^{4}$ instances of the model generated for a wide range of parameters. The dark grey line is the best fit polynomial to the running median of the $10^{4}$ instances. The underlying greyscale histogram represents the relation between $t$ and dark matter fraction for model galaxies in our EAGLE samples. The relationship between $t$ and $f_{\rm{D}}(< 6R_{\rm{d}})$ in our toy model agrees well with that between $t$ and dark matter fraction within $6R_{\rm{d}}$ in model EAGLE galaxies. It is clear that, under the assumption a galaxy is composed of an exponential baryonic disk and a pseudo-isothermal dark halo, the turnover parameter $t$ may be used as a diagnostic tool to predict the dark mass fraction of a galaxy (with the intrinsic uncertainty varying depending on the value of $t$). {\bf Bottom Right:} As for Figure~\ref{fig:gridplot_eagle_obs}, but with blue shading indicating the corresponding median $f_{\rm{D}}(<6R_{\rm{d}})$ according to the toy model. The sub-samples all have values of $t$ corresponding to $f_{\rm{D}}(<6R_{\rm{d}})\gtrsim60$ percent.%
     }%
\label{fig:fdm_vs_turn}
\end{figure*}

There is no strong correlation between an EAGLE model galaxy's stellar mass and whether or not its rotation curve declines at large radii. To investigate whether another intrinsic property of the EAGLE model star-forming galaxies is more correlated with $t$, we also measure the correlations between respectively the projected half-mass size ($R_{50}$), the star-formation rate, and the stellar mass surface density $M_{*}/R_{50}^{2}$ and $t$ of the EAGLE  samples. The latter was motivated by the postulate that declining rotation curves (i.e. low dark matter fractions) may be driven not by an absence of dark matter but by an abundance of baryonic matter. Such an abundance should correlate with an increased measure of stellar surface mass density \citep[e.g][]{Casertano:1991}. 

We find that the rotation curve turnover correlates most strongly with stellar mass surface density in the model galaxies, such that the turnover-surface density correlation exhibits the least scatter at each redshift in EAGLE. This correlation is shown in Figure~\ref{fig:gridplot_eagle_obs}.  The scatter in the correlation is less than that in the stellar mass versus turnover correlation shown in Figure~\ref{fig:massturn_eagle}, implying that the stellar surface mass density is more intrinsically connected to the galaxy rotation curve shape than the stellar mass alone  -- as one might expect if the abundance (or scarcity) of baryons was dominating the shape of the curve. The EAGLE galaxies with the steepest declines (down to $t\approx0.8$) tend to have the highest stellar mass surface densities. These high surface density systems are more prevelant at higher redshift in EAGLE, with the abundance increasing up to $z\approx2$. 

To test whether the stellar mass surface density is a good predictor of the extent to which the rotation curves of {\it observed} galaxies decline, we separate our sample in to bins of redshift and surface density ($M_{*}/R_{\rm{h}}^{2}$) and examine the resultant rotation curve shapes. These curves are presented in Figure~\ref{fig:gridplot_curves_dens}. The results are overlaid onto the EAGLE points in Figure~\ref{fig:gridplot_eagle_obs}. Here we make the implicit assumption that the stellar half-light radii of the observed sample and the stellar half-mass radii of the EAGLE galaxies are an equivalent measurement. We observe a variety of shapes for the rotation curves across the different redshift and density bins for our observed samples. Our measurements generally agree with the trend seen in the EAGLE galaxies, with the vast majority falling within the three-sigma limits of the EAGLE population. Notably though, all of the average curves for the observed galaxies either continue to rise or remain flat out to large radii. 

It is clear that the densities for our observed stacked measurements are not sufficiently high as to expect a significantly declining rotation curve. The stellar mass surface density bin boundaries were chosen as a compromise between covering as wide a range of densities as possible whilst also maintaining sufficient numbers of galaxies in each bin as to reliably measure an average rotation curve. From Figure~\ref{fig:gridplot_eagle_obs} it is clear that model galaxies across all redshifts sampled in EAGLE exhibit a similar median trend between surface density and rotation curve turnover. We also find similar scatter in the relation at each redshift slice in EAGLE. It is thus apparent that, in EAGLE at least, galaxies follow a relation between $t$ and surface density that is largely independent of redshift. We therefore combine galaxies from different redshifts in our {\it observed} sample to measure the higher surface density space of Figure~\ref{fig:gridplot_eagle_obs}, where one would expect to see the strongest decline in galaxy rotation curves (i.e. the lowest values of $t$). 

We produce average rotation curves extracted from the median stacked position-velocity diagrams for $z\gtrsim0.6$ star-forming galaxies in our sample observed with KMOS. These curves are presented in Appendix~\ref{sec:combinedRCs}. The measurements from these curves are shown in Figure~\ref{fig:gridplot_eagle_obs}. As well as a high density bin, we also measure $t$ for stacks from the same combined sample in two lower surface density bins, showing them to generally agree with the measurements made in the individual redshift bins. The average rotation curve associated with the highest density bin possible to robustly measure with our observed sample is consistent with being flat between $3R_{\rm{d}}^{\prime}$ and $6R_{\rm{d}}^{\prime}$. There is thus little evidence to suggest in general that galaxies at high redshift have unusually low central dark matter fractions (within $6R_{\rm{d}}^{\prime}$), including even those systems with higher than average stellar mass surface densities. Most importantly, all of these combined measurements are consistent, within uncertainties, with the trends observed in EAGLE and thus the dark matter properties expected from $\Lambda$CDM theory. 

\subsection{Implied Dark Matter Fractions}
\label{subsec:estimate_fDM}

Finally, we more formally discuss our results in the context of the dark content of star-forming galaxies since $z\approx2.2$. We quantitatively link our turnover parameter, $t$, to the expected dark matter fractions of galaxies and examine the evolution of this quantity with redshift. 

\subsubsection{Rotation Curve Turnover as a Probe of Dark Content}

The shape of the average galaxy rotation curves constructed in this work should be linked to the average dark matter content of the individual galaxies contributing to each average curve. As a means to quantify this dark matter fraction, we draw comparisons with a simple toy model for the rotation velocity such that $v_{mod}^{2}(r) = v_{\rm{disc}}(r)^{2} + v_{2}(r)^{2}$, i.e.\ the baryonic exponential disk plus pseudo-isothermal dark matter halo model described in Equations~\ref{eq:v}--\ref{eq:MH} in Appendix~\ref{sec:RCfittingfunctions}, but where we now do not fix $h=1$. We use the turnover parameter, $t$ to compare this model to the observed curves (where for the intrinsic model there is no requirement to account for beam smearing and thus $t = v_{6R_{\rm{d}}^{\prime}}/v_{3R_{\rm{d}}^{\prime}} \equiv v_{6R_{\rm{d}}}/v_{3R_{\rm{d}}}$).

To translate the $t$ parameter to a dark matter fraction we generate $10^{4}$ instances of our toy model over a wide range of physical parameters. For each instance we calculate the dark mass fraction ($f_{\rm{D}}=M_{\rm{H}}\ /\ M_{\rm{H}} + M_{\rm{d}}$, where $M_{\rm{H}}$ and $M_{\rm{d}}$ are respectively the mass of the halo and the mass of the disk) within $6R_{\rm{d}}$, and also measure $t$ from the output curve. In Figure~\ref{fig:fdm_vs_turn} we show the relationship between $t$ and $f_{\rm{D}}$ according to our toy model. The link between $t$ and $f_{\rm{D}}$ is clear in that the average $t$ increases (albeit non-linearly, and with increasing uncertainty) as a function of increasing dark mass fraction within $6R_{\rm{d}}$. Furthermore, we also show that the toy model trend is in good agreement with that between $t$ and the dark matter fraction within $6R_{\rm{d}}$ for model galaxies in our EAGLE samples. 

As an example of the predictive use of the toy model, in Figure~\ref{fig:fdm_vs_turn} we show the implied dark mass fraction from our toy model for our observed samples as a function of stellar mass surface density. For each of our sub-samples at each redshift, including the combined ($z\gtrsim0.6$) sample in the highest surface mass density bin, the implied dark mass fraction is large ($f_{\rm{D}}(<6R_{\rm{d}}) \gtrsim 60$ percent in all cases).

\subsubsection{Dark Matter Fractions as a Function of Redshift}

To investigate in more detail to what extent the implied average dark matter fraction of star-forming galaxies depends on redshift, and to reduce observational uncertainty from sub-division of our samples by galaxy properties, we return to our earlier measurements of the shapes of the average rotation curves for our {\it total} samples at each redshift (Figure~\ref{fig:total_wrapped_curves}). 

In the toy model, $f_{\rm{D}}$ represents the fraction of the total mass accounted for by the dark matter halo. When considering observed rotation curves, this dark mass fraction will encompasses any mass not well-represented by the disk component in our model. Assuming all baryons in the galaxy reside in a disk with a single scale length (as assumed by our model) then the dark mass fraction is equivalent to the dark {\it matter} fraction (i.e.\ $f_{\rm{D}} \equiv f_{\rm{DM}}$). However, $f_{\rm{D}}$ could also include any baryonic matter present in the galaxy (in particular atomic or molecular gas) with a large-enough scale radius such that it is better described by the halo component of the model within $6R_{\rm{d}}$.  

In Figure~\ref{fig:fdm_vs_turn}, we see that our model EAGLE galaxies follow a trend in $t$-$f_{\rm{DM}}$ that is well-matched by the $t$-$f_{\rm{D}}$ trend for our more simple model. This suggests that $f_{\rm{D}} \equiv f_{\rm{DM}}$ is a reasonable assumption. However, in the higher-redshift Universe there is evidence to suggest that the molecular gas disks of massive, star-forming galaxies are more extended than the stellar disk \citep[e.g.][]{Ivison:2011,Decarli:2016}. Whilst these studies are only based on small samples of galaxies, they suggest that we should at least account for the possibility that significant amounts of gas reside beyond the maximum radii that we are able to probe in this work for our observed samples. In these cases, the dark {\it matter} fraction will be less than the dark {\it mass} fraction (i.e.\ $f_{\rm{DM}} < f_{\rm{D}}$). 

We therefore calculate two measures of $f_{\rm{DM}}$ for our galaxy samples as a function of redshift, considering in turn each of two cases: 1) that all the baryons (stars and gas) in each galaxy in our sample are arranged in a common, single disk with a single scale length (i.e.\ each galaxy obeys the assumption of our toy model that $f_{\rm{D}} \equiv f_{\rm{DM}}$), or 2) the extreme assumption that the distribution of the gas in the galaxies (if present) is sufficiently extended that its entire contribution to the galaxy rotation curve at a given radius is well-described by the halo component of our model (i.e.\ according to our toy model, it behaves like a dark matter halo within $6R_{\rm{d}}$). We proceed with the understanding that the first calculation is the most appropriate of the two and our best estimate of $f_{\rm{DM}}$. Our second calculation simply provides an extreme lower bound to our measure of $f_{\rm{DM}}$ at each redshift. Of course, the ``true'' dark matter fraction may fall somewhere between the two estimates.

For our best estimate, we measure $t$ for the average rotation curve in each (redshift) bin of Figure~\ref{fig:total_wrapped_curves} and calculate the corresponding median average $f_{\rm{D}}(<6R_{\rm{d}})$ for each according to our toy model, simply assuming that $f_{\rm{DM}}=f_{\rm{D}}$, i.e.\ all of the halo mass component is dark matter. 

For our lower estimate, we first determine a molecular gas mass fraction for the galaxies in each redshift bin according to the empirically-motivated prediction of \citet{Sargent:2014} for the molecular gas mass evolution of an average ($5 \times 10^{10} M_{\odot}$ stellar mass) main-sequence star-forming galaxy. \citet{Gobat:2018} show that the measurements of galaxies' molecular gas fractions as a function of redshift from the HERACLES \citep{Leroy:2008}, COLD GASS \citep{Saintonge:2011aa}, PHIBSS \citep{Tacconi:2013}, and EGnoG \citep{Bauermeister:2013} surveys, as well as those from \citet{Daddi:2010} and \citet{Geach:2011}, are all well described by this prediction. We then determine a corresponding estimate of the {\it atomic} gas fraction using the theoretically predicted redshift evolution of the H{\sc i}/H$_{2}$ mass ratio from \citet{ObreschkowRawling:2009}. Combining these for a total (baryonic) gas mass fraction, we then reduce $f_{\rm{D}}$ accordingly to calculate $f_{\rm{DM}}$. 

Our estimates of $f_{\rm{DM}}$ for galaxies in our sample as a function of redshift is shown in Figure~\ref{fig:fdm_vs_z}. We stress here that these are the {\it implied} average dark matter fractions, based on comparison with our toy model and caveat to the validity of the assumptions discussed. The individual estimates are generally consistent with one another given their large uncertainties that encompass the standard error in each of our two calculations described above. At each redshift the dark matter fractions within $6R_{\rm{d}}$ ($\approx13$ kpc) are moderate or large ($f_{\rm{DM}} \geq 67$ percent). 

There is a suggestion that the implied average dark matter fraction for local galaxies is lower than that for the higher redshift population. The estimated $f_{\rm{DM}}$ for the $z\approx0$ galaxies is fairly modest ($f_{\rm{DM}}\approx0.67$). This is in agreement with independent, direct measures of the dark matter fraction ($f_{\rm{DM}} \approx 0.65$--$0.70$, within six disk-scale radii) for 19 THINGS galaxies made by \citet{deBlok:2008}. It is also in agreement with the average dark matter fraction ($f_{\rm{DM}}\approx0.5$) measured within the same radius by \citet{Martinsson:2013b} for 30 local, massive spiral galaxies using their combined H$\alpha$ and H{\sc i} rotation curves. The implied dark matter fraction for galaxies in each of our higher-redshift bins ($z\gtrsim0.6$) is larger. However, given the size of the uncertainties there is no significant difference between our estimates of $f_{\rm{DM}}$ at $z\approx0$ and $z\gtrsim0.6$. Nor is there any significant trend between $f_{\rm{DM}}$ and redshift for our higher-redshift bins. 

\begin{figure}
\begin{minipage}[]{.5\textwidth}
\includegraphics[width=0.95\textwidth,trim= 10 10 10 10,clip=True]{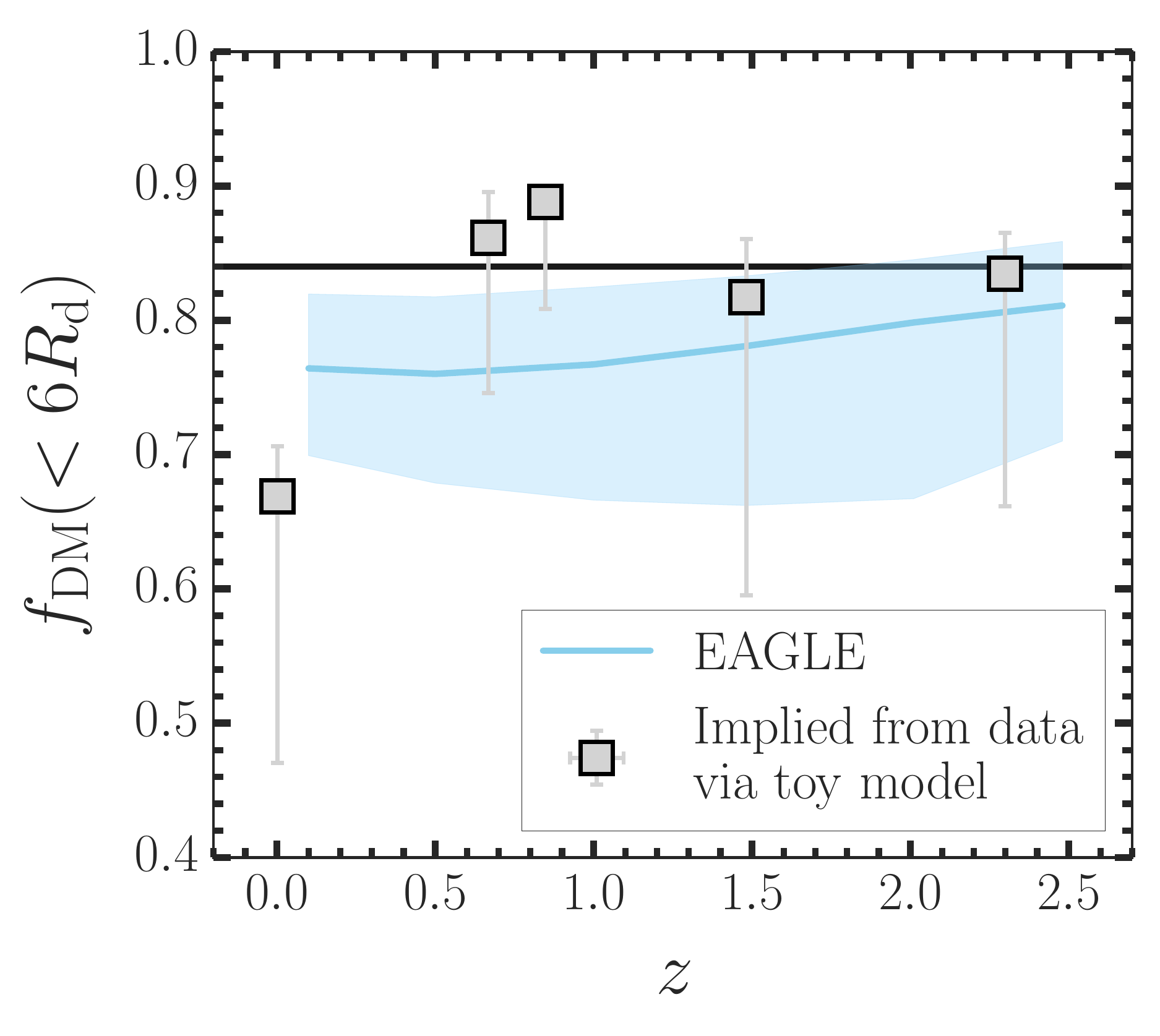}
\end{minipage}
\caption{%
The implied average dark matter fraction ($f_{\rm{DM}}$) within $6R_{\rm{d}}$ as a function of redshift for our samples, estimated from comparison with a baryonic disk plus dark matter halo toy model. Our best estimate assumes that the baryons in each galaxy are present within a single disk with a single scale radius. The lower error bar reflects the possibility that different baryonic components may have different scale radii, and specifically that a substantial amount of gaseous mass in each galaxy is sufficiently extended such that it has the same dynamical effect as a dark halo within $6R_{\rm{d}}$. We find no significant evolution in the implied average dark matter fraction of star-forming galaxies since $z\approx2.2$. The blue solid line and shaded region represent respectively the median and the standard deviation of the dark matter fraction measured within $6R_{\rm{d}}$ for massive ($M_{*}/M_{\odot} > 10^{9}$), star-forming ($>1M_{\odot}\ \rm{yr}^{-1}$) model galaxies in EAGLE (where we linearly interpolate between points to better highlight the general trend). The average trend for model EAGLE galaxies is consistent with our observational, toy model-based estimates. The universal dark matter fraction is indicated with a horizontal line. %
     }%
\label{fig:fdm_vs_z}
\end{figure}

Each of our estimates is consistent with the corresponding measurement for model EAGLE galaxies with $M_{*} > 10^{9} M_{\odot}$ and star-formation rates $>1M_{\odot}\ \rm{yr}^{-1}$; we measure no significant change in their average dark matter fraction as a function of redshift. The lack of significant redshift evolution in $f_{\rm{DM}}$ for both the observed galaxies and model galaxies in EAGLE is perhaps surprising given the prolific rates of star-formation, large gas fractions, short depletion timescales, and high baryonic accretion rates measured for star-forming galaxies at $z\approx1$--$2$ \citep[e.g.][]{Wuyts:2016,Elbaz:2007,Salim:2007,Dutton:2010}. These all imply that large amounts of stellar mass should be assembled in galaxies over the last $\approx10$ Gyr. Thus one might naively expect a large variation in the implied dark matter fraction for our samples as a function of redshift as a result of significant stellar assembly within galaxies.  

We note, however, that there is a clear mass dependence on the redshift evolution of the dark matter fractions of EAGLE model galaxies. Like our observed galaxy sample, the EAGLE sample considered here is dominated in number at each redshift slice by a majority of model galaxies with stellar masses in the range $9 \lesssim \log M_{*}/M_{\odot} \lesssim 10.5$. Within this mass range we see only a small increase in the dark matter fraction within $6R_{\rm{d}}$ with increasing redshift for model EAGLE galaxies. In contrast, for EAGLE galaxies with $\log M_{*}/M_{\odot} \gtrsim 10.5$, there is a much more pronounced redshift evolution in this quantity. Importantly, this evolution is in the opposite direction to the lower mass galaxies; at fixed stellar mass $f_{\rm{DM}}(<6R_{\rm{d}})$ {\it decreases} with increasing redshift. This apparent dichotomy is clearly visible in Figure~\ref{fig:massturn_eagle} for EAGLE. There we see that for EAGLE model galaxies with $\log M_{*}/M_{\odot} \lesssim 10.5$ the median rotation curve turnover, $t$, a proxy for the dark matter fraction, evolves very little with redshift. However, for EAGLE galaxies with $\log M_{*}/M_{\odot} \gtrsim 10.5$, at fixed mass the median $t$ decreases with increasing redshift. This decrease is, however, small ($\Delta t\sim10$ percent in the range $0 \lesssim z \lesssim 2.2$) to the extent that we are unable to determine whether a similar trend exists within our samples of real galaxies given the large uncertainties and observational scatter inherent in our stacked measurements. 

As Figure~\ref{fig:gridplot_eagle_obs} shows, the galaxies with the lowest $t$ are also those with the highest stellar mass surface density. The strong stellar mass dependence in the evolution of $f_{\rm{DM}}(<6R_{\rm{d}})$ in EAGLE can thus be simply explained by considering that the stellar size-mass relation of galaxies evolves with redshift; at fixed stellar mass, galaxies at higher redshift are smaller than those at lower redshift. If there is not a sufficiently corresponding evolution in the size of the dark halo then, within $6R_{\rm{d}}$, the baryons will be effectively concentrated with respect to the dark matter with increasing redshift leading to a corresponding increase in $f_{\rm{DM}}(<6R_{\rm{d}})$. 

In this framework, the mass dichotomy in the redshift evolution of $f_{\rm{DM}}(<6R_{\rm{d}})$ is explained by a similar dichotomy in the strength of the galaxy size-mass evolution. We note that \citet{Furlong:2017} measure a larger increase in galaxy size at fixed stellar mass as a function of redshift for EAGLE galaxies with $\log M_{*}/M_{\odot} \gtrsim 10.5$ than those with lower stellar masses. The authors attribute this to the higher probability for more massive galaxies to undergo merging events \citep{Qu:2017}, expected to be strongly connected to galaxy size growth \citep[e.g.][]{Cole:2000,Naab:2006,vanderWel:2009}. Thus, whilst we find no significant dependence of the implied $f_{\rm{DM}}(<6R_{\rm{d}})$ on redshift for our observed galaxy sample, for galaxies with stellar masses at the higher end of the range spanned by our samples, we expect the dark matter fraction to decrease at fixed mass with increasing redshift.

\section{Conclusions}
\label{sec:conclusions}

In this work we have measured the average rotation curves for thousands of star-forming galaxies at $0 \lesssim z \lesssim 2.2$, tracing the curves out to six stellar disk-scale lengths ($6R_{\rm{d}}^{\prime}$; $\approx$13 kpc, on average). We have shown that the shape of the curves can depend on the scaling prescription used to construct them. Using the least biased (i.e.\ stellar-scaled) prescription, we measured the average rotation curves for galaxies in our sample as a function of redshift, stellar mass, and stellar mass surface density and compared their shape to those of model star-forming galaxies from the EAGLE simulation. Lastly, we provided estimates of the implied dark matter fraction as function of redshift for galaxies in our sample, calculated via comparison with a baryonic disk plus dark matter halo toy model. In this section we summarise our results and present concluding remarks. 

\subsection{The Effects of Different Scaling Prescriptions}
\label{subsec:scalingeffects}

Using stacks of individual rotation curves for galaxies in the range $0.6 \lesssim z \lesssim 2.2$, we showed how the shape of the average galaxy rotation curve depends on the normalisation prescription used to rescale the individual curves in the stack. We demonstrated that:

\begin{itemize}
\item Rescaling the individual curves by their dynamical turnover radius ($R_{\rm{t}}$) measured directly from the curves themselves via the best fitting exponential disk model, and the corresponding observed velocity at that radius ($v_{R_{\rm{t}}}$) i.e.\ adopting the self-scaled scaling prescription of \citet{Lang:2017}, results in an average rotation curve that becomes biased towards smaller, more dispersion dominated systems at large scale radii. This results in a decline in the average rotation velocity with increasing radius. This effect is likely caused or strongly exacerbated by the tendency for the best fitting $R_{\rm{t}}$ to be placed towards the maximum radial extent of the data for flat or rising curves, meaning the majority of the individual curves are compressed to within $\pm1R_{\rm{t}}$.
\smallskip
\item Normalising the individual rotation curves by the stellar disk-scale radius as measured from broadband imaging and modified to account for the seeing ($R_{\rm{d}}^{\prime}$, as described in \S~\ref{subsec:stackRC}), and the velocity at $3R_{\rm{d}}^{\prime}$ ($v_{3R_{\rm{d}}^{\prime}}$) results in an average curve that does not suffer from a strong bias in the sample selection as a function of radius, with the average ratio of rotation-to-dispersion for galaxies contributing to the curve remaining approximately constant with increasing scale radius within the regions probed by the data. 
\end{itemize}

The stellar-scaled normalisation prescription gives an average curve that is more representative of the shape for the typical galaxy in the sample.

\subsection{The Extended Rotation Curves of Star-forming Galaxies}
\label{subsec:RCconclusions} 

Using the stellar-scaled scaling prescription, we stacked deep MUSE and KMOS observations of the spatially-resolved nebular emission from $\approx$1500 star-forming galaxies between $z\approx0.6$--$2.2$ to measure the shapes of their rotation curves at large physical radii. We extracted spectra from each galaxy in bins placed at increasing radii along the galaxy kinematic major axis, constructing a position-velocity diagram for each system normalised in radius, velocity, and flux. We median stacked these position-velocity diagrams in bins of redshift, stellar mass, and stellar mass surface density. Extracting rotation curves from each stacked diagram, we measured the extent to which each curve declined in its outer regions. We quantified this decline using the ``turnover'' parameter $t=v_{6R_{\rm{d}}^{\prime}}/v_{3R_{\rm{d}}^{\prime}}$. We showed that:

\begin{itemize}
\item The average galaxy rotation curves tend to remain flat ($t=1$) or continue to rise ($t>1$) out to $\approx6R_{\rm{d}}^{\prime}$ ($\approx$13 kpc), with an average turnover ranging from $t=1.0$ for galaxies at $z\approx0$ to $t=1.10$--$1.16$ for galaxies in redshift bins $z\approx0.6$--$2.2$.  
\smallskip
\item The extent to which galaxy rotation curves decline trends most strongly with their stellar mass surface density, with those systems with higher surface densities exhibiting flatter or less steeply rising rotation curves between $3R_{\rm{d}}^{\prime}$ and $6R_{\rm{d}}^{\prime}$ (Figure~\ref{fig:gridplot_eagle_obs}). This is in agreement with galaxies in the EAGLE simulation, where we find a correlation between rotation curve shape and stellar surface mass density for massive, star-forming galaxies.
\smallskip
\item The average rotation curve for $z\geq0.6$ galaxies with the highest stellar surface mass densities in our  sample (black points in Figure~\ref{fig:gridplot_eagle_obs}) is consistent with being flat between $3R_{\rm{d}}^{\prime}$ and $6R_{\rm{d}}^{\prime}$ ($t=0.97\pm0.05$). 
\end{itemize}

We argued that the observed trend between galaxies' outer rotation curve slope and surface mass density is likely the manifestation of an increased (or decreased) central baryon density leading to a more negative (or positive) outer slope in the scaled rotation curve rather than any difference in the underlying dark matter density between galaxies. 

\subsection{Implied Dark Matter Fractions}
\label{subsec:fDMconclusions} 

Finally, we estimated the implied dark matter fraction ($f_{\rm{DM}}$) for star-forming galaxies in our sample as a function of redshift by comparing the turnover of their average rotation curve to a simple psuedo-isothermal dark matter halo plus baryonic disk toy model, accounting for the possibility that a large fraction of the gas mass of each galaxy may reside beyond the radii that we probe in this work. 

From comparison with the toy model, the implied average dark matter fraction within $6R_{\rm{d}}$ for our samples of massive, star-forming galaxies is $f_{\rm{DM}}\geq0.67$. This fraction does not significantly change with redshift, and is consistent with the corresponding measurements for massive ($M_{*}/M_{\odot} > 10^{9}$), star-forming ($>1M_{\odot}\ \rm{yr}^{-1}$)  galaxies in the EAGLE simulation. 

We noted that, according to EAGLE, we expect some redshift dependence for the dark matter fractions of galaxies more massive than the majority of those in our sample ($\log M_{*}/M_{\odot} \gtrsim 10.5$). These systems are expected to exhibit a much stronger size evolution at fixed mass over time due to an increased frequency of merging events in comparison to lower mass galaxies.

In summary, we find no observational evidence for any significant deviation from the rotation curve shapes expected based on $\Lambda$CDM theory, observing the average rotation curves for massive, star-forming galaxies at $0\lesssim z \lesssim 2.2$ to either remain flat or continue to rise out to large scale radii. Accordingly, under reasonable assumptions the implication is that these galaxies have correspondingly large dark matter fractions.

\section*{Acknowledgments}

We thank the anonymous referee for their constructive report on this work which helped us to significantly improve its content and clarity.

ALT, AMS, and IRS acknowledge support from STFC (ST/L00075X/1 and ST/P000541/1). ALT also acknowledges support from the ASTRO 3D Visitor program and the ERC Advanced Grant DUSTYGAL (321334). IRS also acknowledges support from the ERC Advanced Grant DUSTYGAL (321334) and a Royal Society/Wolfson Merit Award. MS is supported by VENI grant 639.041.749. SG acknowledges the support of the Science and Technology Facilities Council (ST/L00075X/1). GEM acknowledges support from the Carlsberg Foundation, the ERC Consolidator Grant funding scheme (project ConTExt, grant number No. 648179), and a research grant (13160) from Villum Fonden. PNB acknowledges support from grant ST/M001229/1.

We thank the FMOS-COSMOS team for their invaluable contributions to the KGES target selection. We thank Fabian Walter for providing advice on the THINGS sample. We thank Luca Cortese, Danail Obreschkow, Karl Glazebrook, Edo Ibar, and Omar Almaini for useful discussions on the content of this work. We acknowledge the Virgo Consortium for making their simulation data available. The EAGLE simulations were performed using the DiRAC-2 facility at Durham, managed by the ICC, and the PRACE facility Curie based in France at TGCC, CEA, Bruyres-le-Chtel. 

Based on observations made with ESO Telescopes at the La Silla Paranal Observatory under the programme IDs 60.A-9460, 092.B- 0538, 093.B-0106, 094.B-0061, 095.A-0748, 095.B-0035, 096.A-0200, 097.A-0182, 098.A-0311, and 0100.A-0134. This work is based in part on data obtained as part of the UKIRT Infrared Deep Sky Survey. This work is based on observations taken by the CANDELS Multi-Cycle Treasury Program with the NASA/ESA HST, which is operated by the Association of Universities for Research in Astronomy, Inc., under NASA contract NAS5-26555. HST data was also obtained from the data archive at the Space Telescope Science Institute.




\bibliographystyle{mnras}
\bibliography{tiley_curve_stack.bib} 




\appendix

\section{Rotation Curves: Further Checks}
\label{sec:appendixchecks}

\subsection{The effect of seeing on the stellar-scaled rotation curve shape}
\label{subsec:seeingeffect}

\begin{figure}
\begin{minipage}[]{.5\textwidth}
\includegraphics[width=.95\textwidth,trim= 20 10 5 5,clip=True]{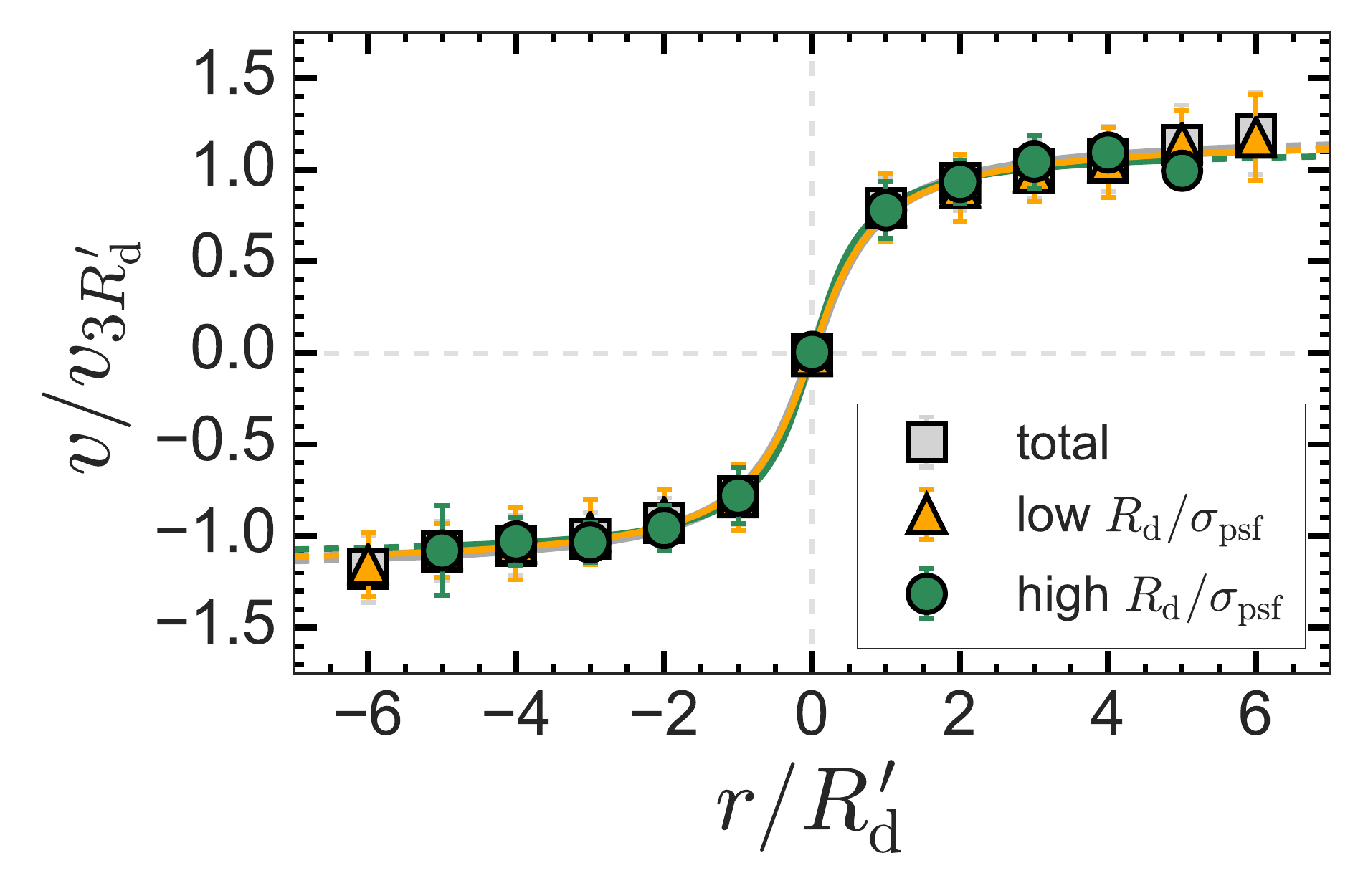}
\end{minipage}
\caption{%
A comparison between the median-average $z\approx0.9$ (KROSS) stellar-scaled rotation curve constructed from individual galaxy rotation curves for respectively the total KROSS sample, those KROSS galaxies that are smallest compared to the size of the PSF (defined as those with a ratio $R_{\rm{d}}/\sigma_{\rm{psf}}$ lower than the 30$^{\rm{th}}$ percentile of the total distribution for KROSS galaxies), and those KROSS galaxies that are largest compared to the size of the PSF (defined as those with a ratio $R_{\rm{d}}/\sigma_{\rm{psf}}$ larger than the 60$^{\rm{th}}$ percentile of the total distribution for KROSS galaxies). The corresponding best fit model (see \S~\ref{subsec:modelstack}) to each rotation curve is also displayed as a solid line (dashed where extrapolated beyond the measurements). The shapes of the three curves are in good agreement with no significant deviations between the three. The ratio of the galaxy size to the size of the PSF has no significant impact on the shape of the average stellar-scaled KROSS rotation curve. %
     }%
\label{fig:seeingeffects}
\end{figure}

In Figure~\ref{fig:seeingeffects} we compare the shape of the median-average of the individual stellar-scaled KROSS rotation curves in two bins of $R_{\rm{d}}/\sigma_{\rm{psf}}$. We measure the average curve for galaxies in respectively the lower 30$^{\rm{th}}$ percentile and upper 60$^{\rm{th}}$ percentile in $R_{\rm{d}}/\sigma_{\rm{psf}}$ for the KROSS sample. The bin positions were chosen in order to produce the maximum possible difference in the average $R_{\rm{d}}/\sigma_{\rm{psf}}$ between the galaxies contributing to each of the two curves, whilst simultaneously maintaining a reasonable (sub-)sample size in each case for an accurate measure of the rotation curve and, in the case of the upper $R_{\rm{d}}/\sigma_{\rm{psf}}$ bin, sufficient radial extent of the data to measure the rotation curve out to $\approx6R_{\rm{d}}^{\prime}$. We also compare these two curves to that for the total KROSS sample. 

Figure~\ref{fig:seeingeffects} shows that the ratio of the galaxy size to the size of the PSF makes no significant difference to the shape of the median-average stellar-scaled KROSS rotation curve within $\pm 6R_{\rm{d}}^{\prime}$, with the shapes of all three curves in good agreement. We find similar results for the other galaxy samples, implying that beam smearing does not significantly affect the shape of the average stellar-scaled rotation curves presented in this work. 

\subsection{Alternative emission line fitting and its effect on rotation curve shape}
\label{subsec:altfittingeffect}

Beam-smearing effects in integral field spectroscopy observations can skew the observed profiles of emission lines in the data cubes, causing the lines to become asymmetric (non-Gaussian) with broad tails of emission towards the systemic galaxy velocity. It is difficult to judge a priori to what extent this effect is present in the data used in this work and, most importantly, whether or not it has a significant impact on the shapes of the average rotation curves we construct. 

In this sub-section we therefore use the KGES ($z\approx1.5$) KMOS data cubes as a test set in order to determine the extent of the asymmetry of the nebular emission lines in our data. We re-fit all the H$\alpha$ and [N {\sc ii}] emission lines in each spaxel of each of the KGES datacubes using a Gaussian-hermite model, i.e.\ the same triple Gaussian model described in \S~\ref{subsec:linemaps}, but now including an $h_{3}$ term (where $h_{3}$ describes asymmetric deviations from a Gaussian, and is tied for each of the H-alpha, and two [N {\sc ii}] lines) that is free to vary between $-0.3 < h_{3} < 0.3$. The limits represent the points beyond which the model profile becomes highly unphysical. 

For every datacube (i.e. for each galaxy) we compared the observed line-of-sight velocity in each spaxel from both the triple Gaussian fit (\S~\ref{subsec:linemaps}) and the triple Gaussian-hermite fit. The velocity residuals ($\Delta v$) are shown in Figure~\ref{fig:delta_v_Gauss_Gaussherm} as a function of both the H$\alpha$ $S/N$ and best fit $h_{3}$ value. In general, the velocities from the two fits disagree when the absolute value of $h_{3}$ is highest. However, this also preferentially occurs at low H$\alpha$ $S/N$, when the addition of the $h_{3}$ term is not statistically warranted. Figure~\ref{fig:delta_v_Gauss_Gaussherm} also shows that there is no evidence for a systematic offset between the two measures of velocity (the mean is $-0.3 \pm 0.1$ km s$^{-1}$) and the scatter too is small (the standard deviation is $13$ km s$^{-1}$). We conclude that the two measures of velocity are in good agreement and therefore that the nebular emission lines in our data are not skewed as a result of beam smearing effects to such an extent that our triple Gaussian model is an inaccurate description of their shape.

\begin{figure*}
\begin{minipage}[]{1.\textwidth}
\centering
\includegraphics[width=.95\textwidth,trim= 20 20 10 10,clip=True]{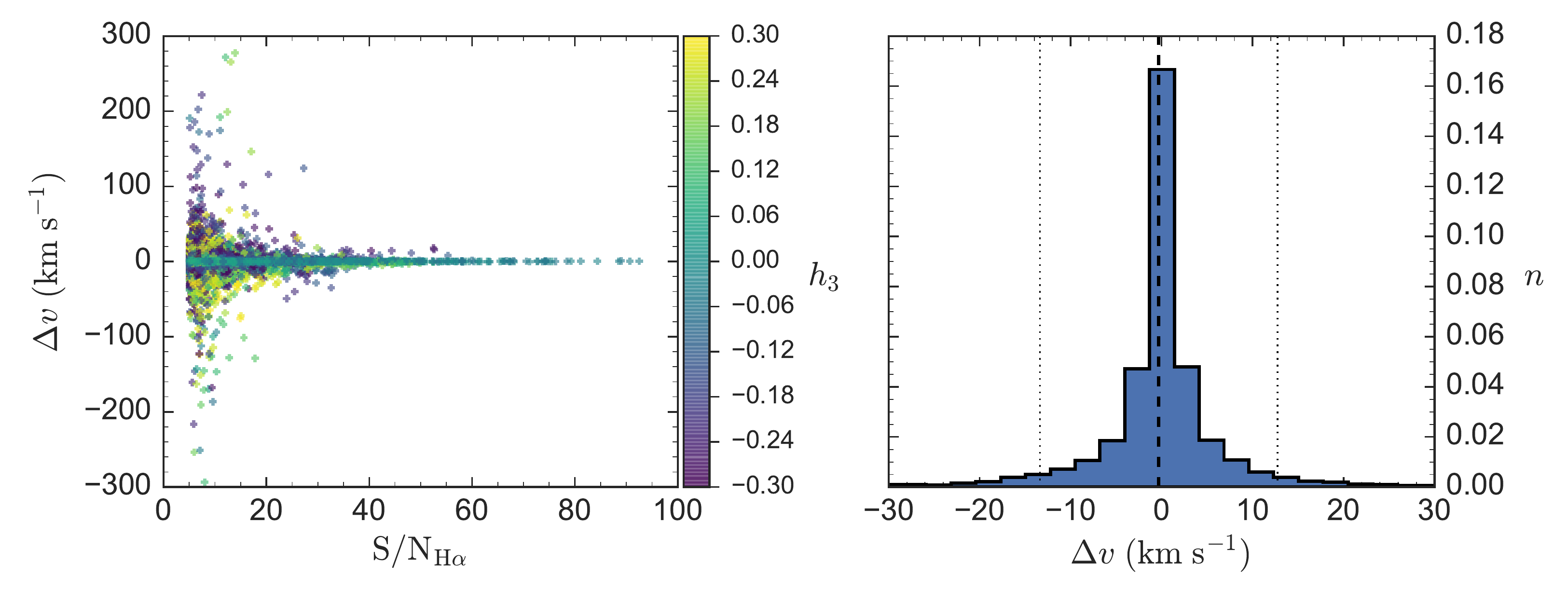}
\end{minipage}
\caption{%
{\bf Left:} The residuals between the observed line-of-sight velocities derived from Gaussian triplet fits and Gaussian-hermite fits (i.e. including an $h_{3}$ term) for every spaxel of the $z\approx1.5$ sample KMOS data cubes as a function of H$\alpha$ $S/N$ ($S/N_{\rm{H}\alpha}$), where both the Gaussian and Gaussian-hermite fit have $S/N_{\rm{H}\alpha} \geq 5$. The points are colour-coded by the $h_{3}$ value in the best fit Gaussian-hermite triplet. The residuals are largest at low $S/N_{\rm{H}\alpha}$ and high values of $h_{3}$. {\bf Right:} The (normalised) distribution of the velocity residuals between the Gaussian and Gaussian-hermit fits for each spaxel of the $z\approx1.5$ data cubes. There is no evidence for strong differences between the two velocity measures, with the mean and standard deviation of the distribution as $-0.3 \pm 0.1$ km s$^{-1}$ and $13$ km s$^{-1}$, respectively.%
     }%
\label{fig:delta_v_Gauss_Gaussherm}
\end{figure*}

Of course, from Figure~\ref{fig:delta_v_Gauss_Gaussherm} alone we cannot tell how any differences between the two measures of rotation velocity (Gaussian versus Gaussian-hermite) propagate into the individual rotation curves and thus the final average of the (scaled) curves. To test how this might affect our final conclusions, we construct median average stellar-scaled rotation curves for the KGES galaxies using both measures, via the same methods described in \S~\ref{subsec:stackRC} of the paper. For this test we only consider those galaxies with sufficient pixes in both the Gaussian- and Gaussian-hermite-derived velocity maps to reliably measure a rotation velocity from both. The two average curves are shown in Figure~\ref{fig:RC_comp_Gauss_Gaussherm}. They are in good agreement. We therefore find no evidence that the shapes of the average rotation curves presented in \S~\ref{subsec:stackRC} are significantly affected by the use of a Gaussian triplet model to describe the H-alpha and [N {\sc ii}] emission, rather than a triple Gaussian-hermite model. 

\begin{figure}
\begin{minipage}[]{.5\textwidth}
\includegraphics[width=.95\textwidth,trim= 23 10 10 20,clip=True]{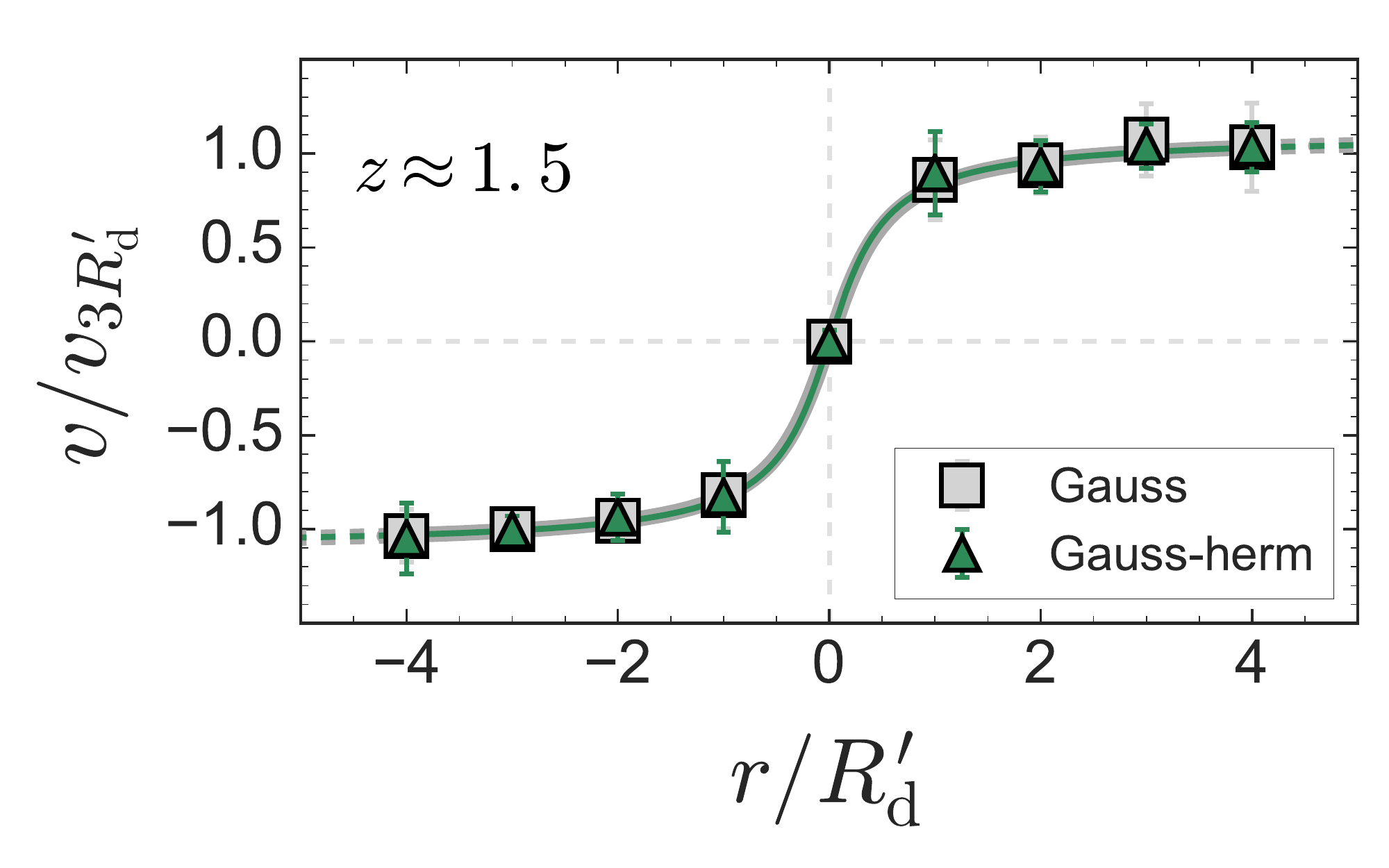}
\end{minipage}
\caption{%
The median-average of the individual stellar-scaled rotation curves for our $z\approx1.5$ galaxy sample, extracted from velocity maps derived from either Gaussian triplet or Gaussian-hermite triplet fits to the emission lines in each of the data cubes. The average curves include those galaxies with sufficient pixes in both the Gaussian- and Gaussian-hermite-derived velocity maps to reliably measure a rotation velocity from both. The two curves are in good agreement, with no significant deviations in the range $\pm4R_{\rm{d}}^{\prime}$.%
     }%
\label{fig:RC_comp_Gauss_Gaussherm}
\end{figure}

\subsection{The effect of spatial binning on rotation curve shape}
\label{subsec:binningeffect}

\begin{figure*}
\begin{minipage}[]{1.\textwidth}
\includegraphics[width=.95\textwidth,trim= 0 15 10 10,clip=True]{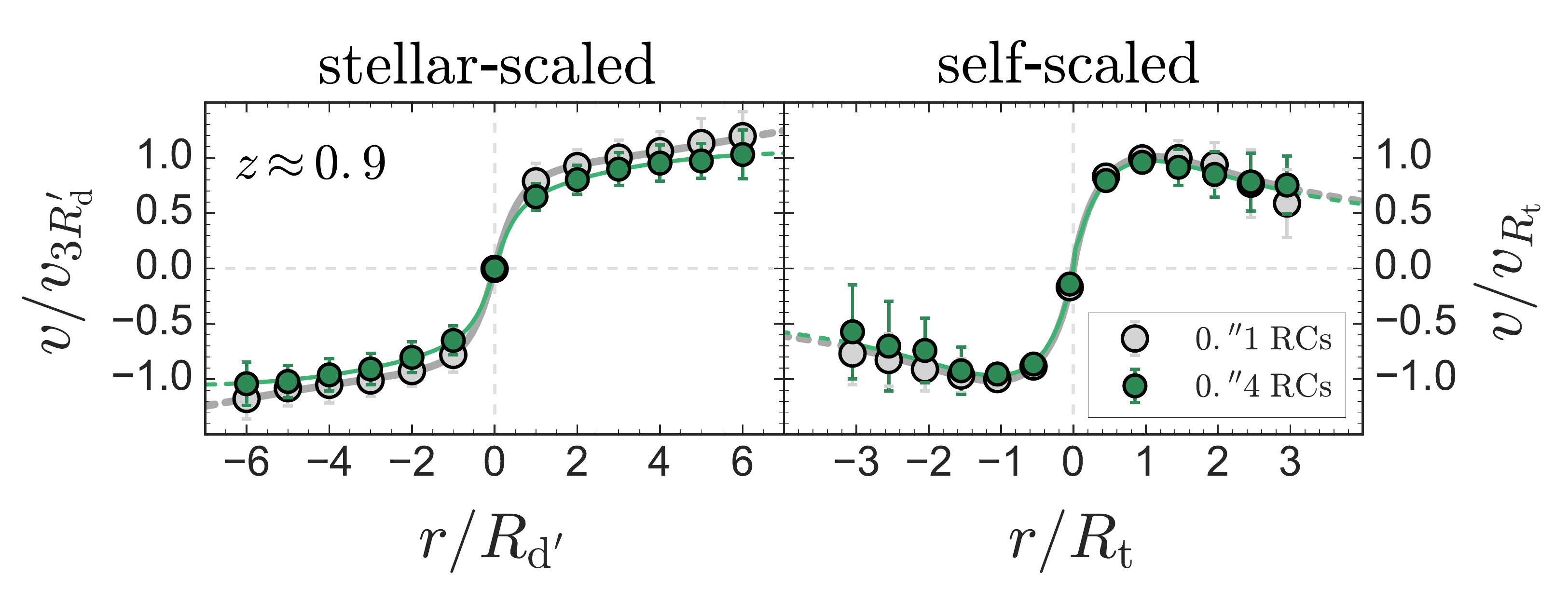}
\end{minipage}
\caption{%
The median-average $z\approx0.9$ (KROSS) stellar-scaled rotation curve (left panel) and the corresponding self-scaled curve (right panel) constructed from individual galaxy rotation curves extracted from velocity maps (as described in \S~\ref{subsec:linemaps}) sampled with either $0\farcs1$ or $0\farcs4$ spaxels. The two different samplings result in curves that are in good agreement, within uncertainties, although the average of the $0\farcs4$-sampled stellar-scaled curves is slightly shallower than the corresponding $0\farcs1$-sampled average curve. Most importantly, for either choice of spatial sampling, we draw the same conclusion - that the shape of the median average rotation curve {\it for the same galaxies} starkly differs depending on the normalisation technique.%
     }%
\label{fig:binningeffects}
\end{figure*}

In Figure~\ref{fig:binningeffects} we demonstrate that the spatial sampling of the galaxy velocity maps makes no significant difference to the shape of the median average of the rotation curves extracted from them, and thus that the result presented in Figure~\ref{fig:medRCstacks} and discussed in \S~\ref{subsec:stackRC} (i.e. that the shape of the average of the individual galaxy rotation curves is strongly dependendent on the normalisation technique used to construct it) is robust to our initial choice of sampling.

\subsection{Stacked rotation curves versus stacked position-velocity diagrams}
\label{subsec:pvversusRS}
  
In this work we present average rotation curves constructed either by taking a median average of individual galaxy rotation curves (\S~\ref{subsec:stackRC}), or by median-averaging individual galaxy position-velocity diagrams (\S~\ref{subsec:specs}). In this sub-section we verify that the shape of the average rotation curve does not differ as a function of the method of construction. We compare the shape of the average rotation curve extracted from the stacked KROSS ($z\approx0.9$) position-velocity diagrams to that of the median of the individual rotation curves for the same galaxies. In Figure~\ref{fig:rotcurvmethodcheck}, we show that the rotation curves derived using either method agree well, within uncertainties, verifying that the shape of the average rotation curve is not dependent on its method of construction. We find similar results for galaxies in our sample at other redshifts.
  
\begin{figure}
\begin{minipage}[]{.5\textwidth}
\includegraphics[width=.95\textwidth,trim= 20 15 10 10,clip=True]{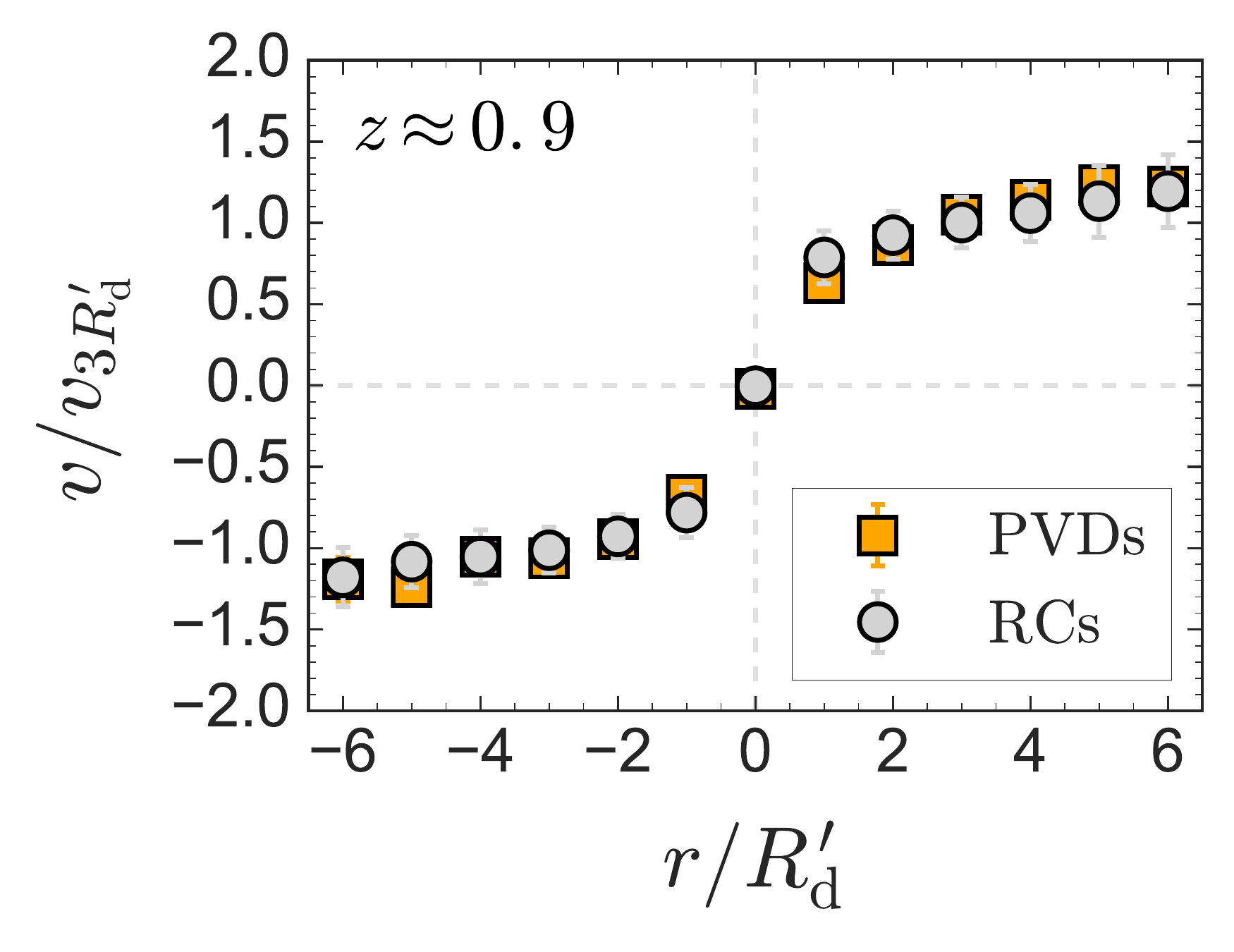}
\end{minipage}
\caption{%
A comparison between the median-average of the individual $z\approx0.9$ (KROSS) stellar-scaled rotation curves (grey points; see \S~\ref{subsec:stackRC}) and the corresponding rotation curve for the same galaxies derived from stacked galaxy H$\alpha$ emission using the method outlined in \S~\ref{subsec:specs}. The two curves are in good agreement, within uncertainties (despite the latter rising slightly more slowly in its inner regions than the former)  thus verifying the validity of our methods. %
     }%
\label{fig:rotcurvmethodcheck}
\end{figure}

\subsection{Recovering a turnover}
\label{subsec:recoveraturn}

In this sub-section we demonstrate that adopting the scaling prescription of \cite{Lang:2017} leads to us recovering an average galaxy rotation curve from the median stack of individual galaxy position-velocity diagrams that declines in its outer regions to the same extent as the average curves presented in \citet{Lang:2017}. 

In Figure~\ref{fig:turncheck} we present average rotation curves for the combined $z\approx1.5$ and $z\approx2.2$ KMOS samples presented in this work. We select these galaxies to match as closely as possible the redshift for the stacking sample in \citet{Lang:2017}, whilst also maintaining a reasonable number of galaxies to maintain as strong a signal as possible when stacking their nebular emission. In the right panel of the figure, we present the rotation curve derived from stacked position-velocity diagrams in the manner described in \S~\ref{subsec:specs} but normalising the position-velocity diagrams in radius and velocity using respectively the turnover radius ($R_{\rm{t}}$, as determined from the best fit exponential disc model (Equation~\ref{eq:mod}) to the individual galaxy rotation curve), and the maximum observed rotation velocity ($v_{\rm{max}}$) to match the adopted prescription of \citeauthor{Lang:2017}. We also include the average rotation curve from Figure~5 of that work. The two curves follow the same shape, exhibiting identical declines at large fractional radii. In the left panel of the figure we demonstrate that, for the same galaxies (i.e.\ our combined $z\approx1.5$ and $z\approx2.2$ KMOS sample), the stellar-scaled rotation curve remains flat out to large scale radii. This is in agreement with our findings presented in \S~\ref{subsec:stackRC}, that the shape of the rotation curve is dependent on the scaling prescription adopted in its construction. 

We conclude that the decline in the self-scaled curve is driven by the scaling prescription and is not a robust measure of the rotation curve shape for {\it typical} star-forming galaxies at high redshifts.

\begin{figure*}
\centering
\begin{minipage}[]{1.\textwidth}
\centering
\includegraphics[width=.95\textwidth,trim= 20 10 20 10,clip=True]{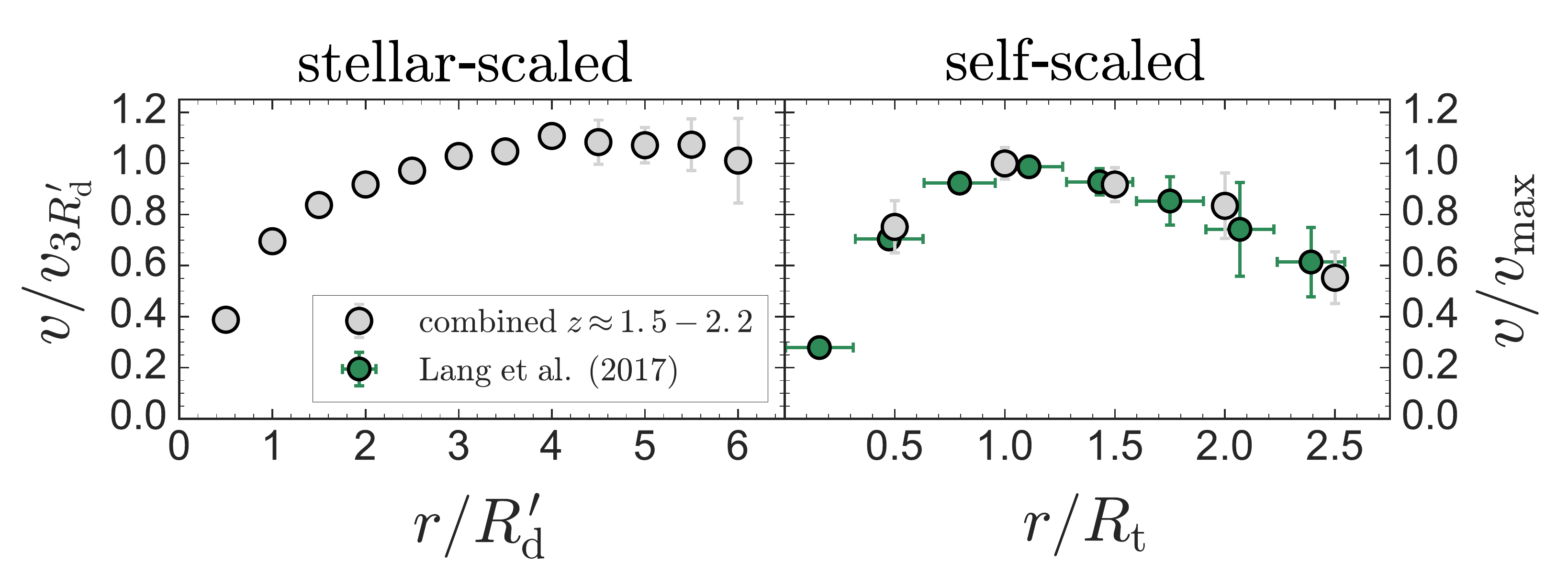}
\end{minipage}
\caption{%
A demonstration of our ability to recover a declining rotation curve from stacked position-velocity diagrams adopting the self-scaled ($R_{\rm{t}}$,$v_{\rm{max}}$) normalisation prescription of \citet{Lang:2017}. The grey points represent the median rotation curve extracted from the stacked position-velocity diagrams of the combined $z\approx1.5$ and $z\approx2.2$ KMOS samples. This combined sub-sample is designed to be large enough to reduce the effects of noise in the resultant average rotation curve whilst also comprising those galaxies best matched in redshift to the stacking sample in \citet{Lang:2017}. The median stellar mass of the combined sample is $\log M_{*}/M_{\odot} = 10.3 \pm 0.3$, where the uncertainty is the median absolute deviation from the median. The green points represent the average rotation curve as constructed by \citet{Lang:2017} and presented in Figure~5 of that work. {\bf Left: } The {\it stellar-scaled} rotation curve for the combined $z\approx1.5$ and $z\approx2.2$ KMOS samples {\it remains flat}. {\bf Right:} The two {\it self-scaled} curves exhibit the same shape, including a strong decline in velocity at large fractional radii. For the same galaxies (i.e.\ the grey points in each panel), the rotation curve either remains flat (stellar-scaled) or turns over (self-scaled), depending on the normalisation prescription.%
     }%
\label{fig:turncheck}
\end{figure*}

\subsection{Position-velocity diagrams after binning within a redshift slice}
\label{subsec:morebinpvdiags}

In this sub-section we verify that we are able to extract rotation curves from stacks of individual position-velocity diagrams from our star-forming galaxies after simultaneously binning the galaxies by two parameters. In Figure~\ref{fig:gridplot} we present the position-velocity diagrams for our galaxy sample binned in both redshift and and stellar mass. At first inspection there is no obvious decline in the rotation velocity with increasing radius out to $\approx6R_{\rm{d}}^{\prime}$ for the majority of the rotation curves, regardless of redshift or stellar mass. The exception is the lowest stellar mass bin for the $z\approx1.5$ row, for which the rotation curve does appear to significantly decline beyond its peak. However, this decline is not symmetrical (one side of the curve declines much more strongly than the other) and it is clear that the signal is weaker in this particular position-velocity diagram. To address this, before considering the shapes of the average curves in detail, in the main text we boost the signal in the average position-velocity diagram by first folding the individual galaxy position-velocity diagrams about their own axes before stacking them.

\begin{figure*}
\centering
\begin{minipage}[]{1\textwidth}
\centering
\includegraphics[width=1.\textwidth,trim= 20 10 45 30,clip=True]{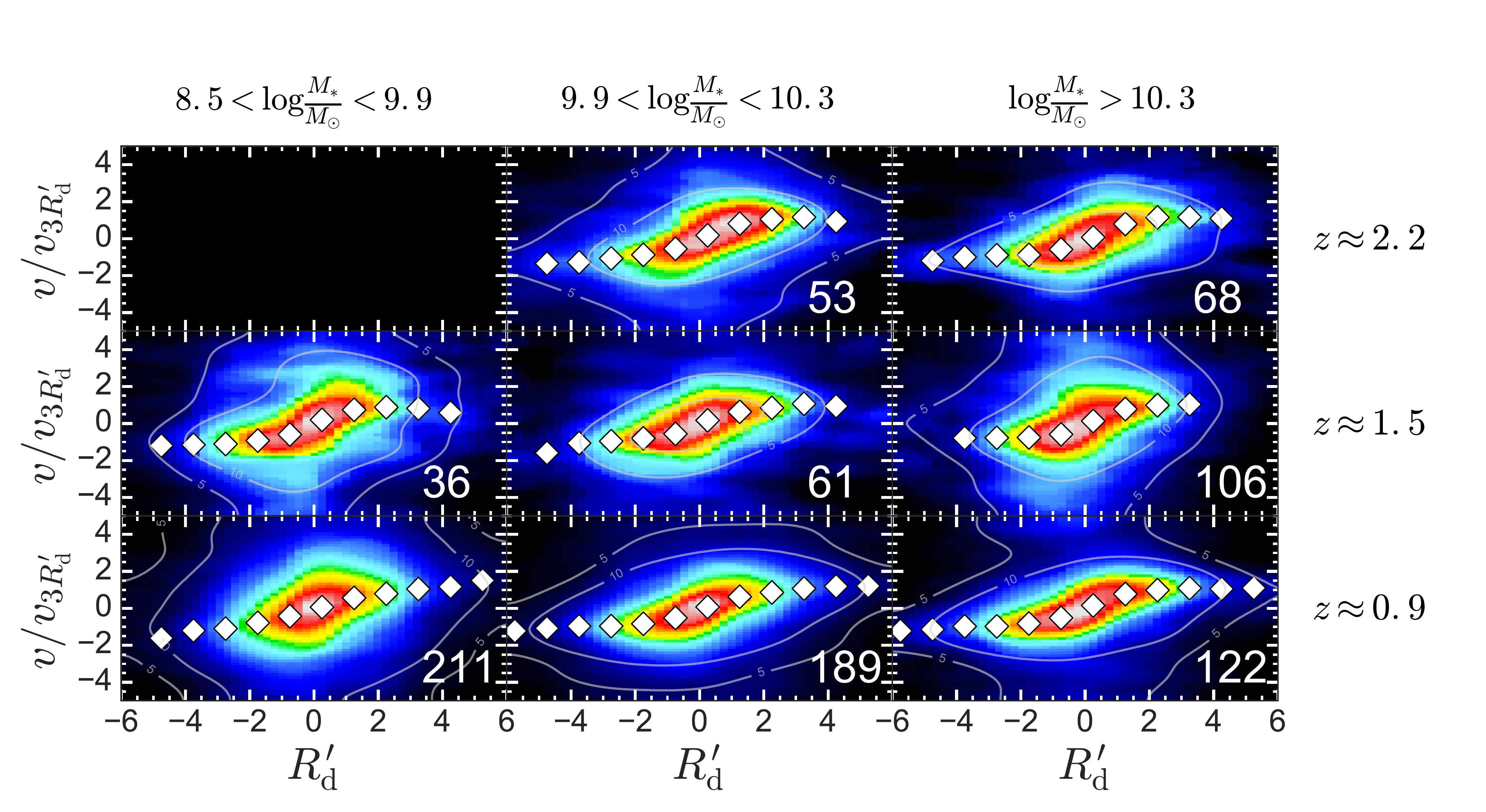}
\end{minipage}
\caption{%
Median-averaged normalised position-velocity diagrams for samples of galaxies from our analysis separated in to bins of redshift and mass. The position-velocity diagrams are derived from stacked H$\alpha$ emission. We overlay grey contours corresponding to a signal-to-noise ratio of 5 and 10. The white diamonds in each panel represent the peak position of the best fit to the spectrum corresponding that radial column. The majority of the curves remain flat or continue to rise out to large radii.%
     }%
\label{fig:gridplot}
\end{figure*}

\section{The relationship between galaxy size, S\'ersic index, and rotation-dominance}
\label{sec:stackbias}

In this section we examine the extent to which the size, and the S\'ersic index, of a galaxy correlates with the dominance of rotation or dispersion in the galaxy's kinematics. In the top panel of Figure~\ref{fig:vonsigvsr} we show that the median size of dispersion-dominated galaxies ($v_{\rm{c}}/\sigma_{0} \lesssim 1$) is lower than for rotation-dominated galaxies. This trend is emphasised when galaxy size is measured using the dynamical turnover radius from the best fit exponential disk model the galaxy rotation curve. This suggests that $v_{\rm{c}}/\sigma_{0} \lesssim 1$ galaxies may also be those systems that are more compact. This has important implications for the choice of scaling prescription used to normalise galaxy rotation curves, as discussed in \S~\ref{subsubsec:normbiases}. In particular it demonstrates the bias toward low $v_{\rm{c}}/\sigma_{0}$ systems at larger radii in the self-scaled curves discussed in that section is likely partly a symptom of selecting for systems with small $R_{\rm{t}}$.

The bottom panel of Figure~\ref{fig:vonsigvsr} presents the median S\'ersic index ($n_{\rm{s}}$; as measured from $H$-band {\it HST} imaging by \citealt{vanderWel:2012}), available for 102 KROSS galaxies in CANDELS, as a function of median $v_{\rm{c}}/\sigma_{0}$. It shows that dispersion-dominated galaxies ($v_{\rm{c}}/\sigma_{0} \lesssim 1$) have a higher median average S\'ersic index than that for rotation-dominated galaxies ($v_{\rm{c}}/\sigma_{0} \gtrsim 1$) or for the total sub-sample. 

This median value has comparatively large uncertainty, due to the small number of galaxies per bin as only a sub-set of KROSS galaxies fall within the CANDELS fields. Nevertheless, this trend and the corresponding decrease in size for dispersion-dominated galaxies shown in the top panel of Figure~\ref{fig:vonsigvsr} together suggest that these small, dispersion-dominated systems with above average S\'ersic indices, that dominate the outskirts of the self-scaled average rotation curves presented in Figures~\ref{fig:medRCstacks} and \ref{fig:biascheck}, are in fact less disk-like than those larger, more-rotation dominated systems contributing to the inner regions of the self-scaled curves. 

If true, this would suggest that the decline in the average, self-scaled rotation curve seen in the stacked curves may actually be the result of a conspiracy between a more prominent (pseudo-)bulge component \citep[pseudo-bulges should have $n\lesssim2$ e.g.][]{Fisher:2008} in the individual galaxy rotation curves and the application of a common radial scaling of each curve, meaning only a region of the curve that is, by definition, Keplarian in its shape is considered.

\begin{figure}
\begin{minipage}[]{.5\textwidth}
\includegraphics[width=.95\textwidth,trim= 10 15 10 0,clip=True]{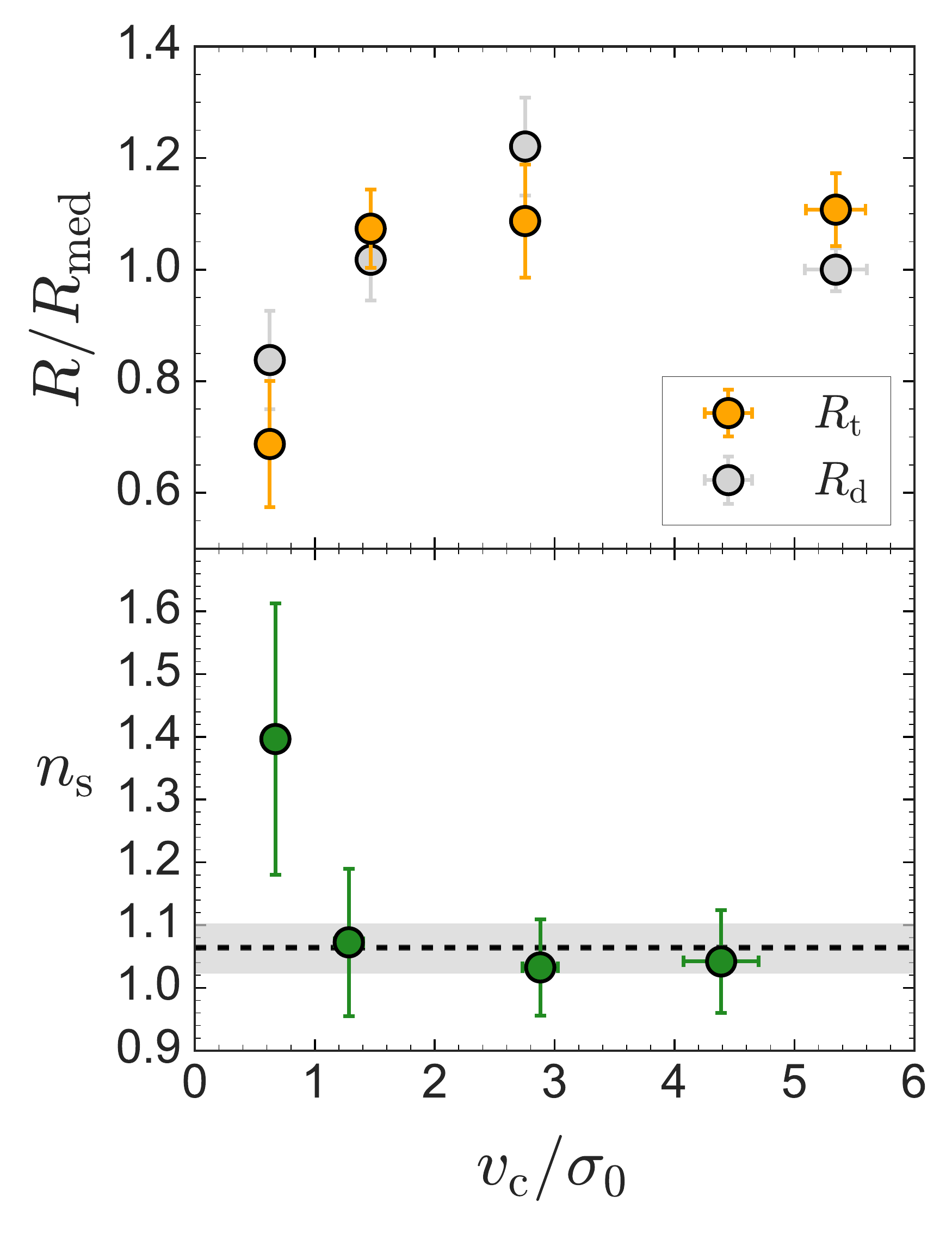}
\end{minipage}
\caption{%
{\bf Top:} The median size of the KROSS galaxies as a function of $v_{\rm{c}}/\sigma_{0}$ and with respect to the overall median size of the sample. We include the trends for both the dynamical turnover radius measured from the galaxy rotation curves ($R_{\rm{t}}$), and the stellar disk-scale radius ($R_{\rm{d}}$), measured from broadband imaging. Both measures of size remain approximately constant within uncertainties for $v_{\rm{c}}/\sigma_{0} \gtrsim 1$, but  decrease with respect to the overall median of the sample for $v_{\rm{c}}/\sigma_{0} \lesssim 1$. This effect is seen most strongly for the $R_{\rm{t}}$ size measurement. {\bf Bottom:}  The median S\'ersic index (as measured from $H$-band {\it HST} imaging by \citealt{vanderWel:2012}) as a function of median $v_{\rm{c}}/\sigma_{0}$ for a sub-set of 102 KROSS galaxies within the CANDELS fields. The black dashed line and grey shaded region represent respectively the median S\'ersic index for the total sample and its standard uncertainty. Dispersion-dominated systems ($v_{\rm{c}}/\sigma_{0} \lesssim 1$) have a larger median S\'ersic index (despite a large associated uncertainty) than that for rotation-dominated systems ($v_{\rm{c}}/\sigma_{0} \gtrsim 1$) or for the total sub-set.%
     }%
\label{fig:vonsigvsr}
\end{figure}

\section{Rotation Curve Fitting Functions}
\label{sec:RCfittingfunctions}

To trace the intrinsic shape of the normalised, average rotation curves presented in this work, we find the best fit to the data for three commonly employed models. First, we fit the arctangent disk model for galaxy rotation curves \citep{Courteau:1997aa} that gives the rotation velocity as a function of radius as

\begin{equation}\label{eq:arctan}
v(r)=\frac{2}{\pi}v_{\rm{max}}\arctan\frac{r}{r_{\rm{dyn}}}\,\,,
\end{equation} 

\noindent where $v_{\rm{max}}$ is the rotation velocity at infinite radius, and $r_{\rm{dyn}}$ is the characteristic radius associated with the arctangent turnover. Second we fit the exponential disk model described in Equation~\ref{eq:mod} (where we now fix $v_{\rm{off}}=0$). Finally, we fit a model with contributions from a scaled, exponential disk (normalised in radius by the disk scale length and in velocity by its maximum) and a pseudo-isothermal dark matter halo such that the rotation velocity as a function of radius $v^{2}(r) = v_{1}^{2}(r) + v_{2}^{2}(r)$, where  

\begin{equation}\label{eq:v}
v_{\rm{disc}}(r)^{2}=\frac{r^2 \pi G \mu_{0}}{h}(I_{0}K_{0}-I_{1}K_{1})\,\,,
\end{equation}

\noindent and

\begin{equation}\label{eq:v1}
v_{1}(r)=A \times v_{\rm{disc}}\,\,,
\end{equation}

\noindent where $A$ is a scaling factor. The shape of the rotation curve for a purely exponential disk will always take exactly the same form when normalised in radius and velocity. Since we fit the model to observed rotation curves that are themselves normalised, we therefore set $h=1$ (since $r$ is in units of $h$ in the normalised curves) to fix its shape but allow it to vary in amplitude. Any deviation from this shape seen in the observed data must then be described by the halo component of the model. In this manner we prevent the disk component in the best fit from accounting for any aspect of the shape of the observed normalised rotation curve that can in reality only be due to the presence of a second, non-exponential disk (i.e.\ halo) component. 

The halo rotation velocity component is given as 

\begin{equation}\label{eq:v2}
v_{2}^{2}(r) = GM_{\rm{h}}(r)/r\,\,.
\end{equation}

\noindent  The halo mass within $r$ is given as

\begin{equation}\label{eq:MH}
M_{\rm{h}}(r) = \int \frac{4 \pi p_{0} r^{2}}{1+(r/r_{0})^{\alpha}}\mathrm{d}r\,\,,
\end{equation}

\noindent where $p_{0}$ is the central halo density, $r_{0}$ is the halo scale radius, and $\alpha=2$ is the slope of the density profile at large radii.

\section{Rotation Curves for a Combined Higher-Redshift Sample}
\label{sec:combinedRCs}

\begin{figure*}
\centering
\begin{minipage}[]{1.\textwidth}
\centering
\includegraphics[width=1.\textwidth,trim= 0 0 0 0,clip=True]{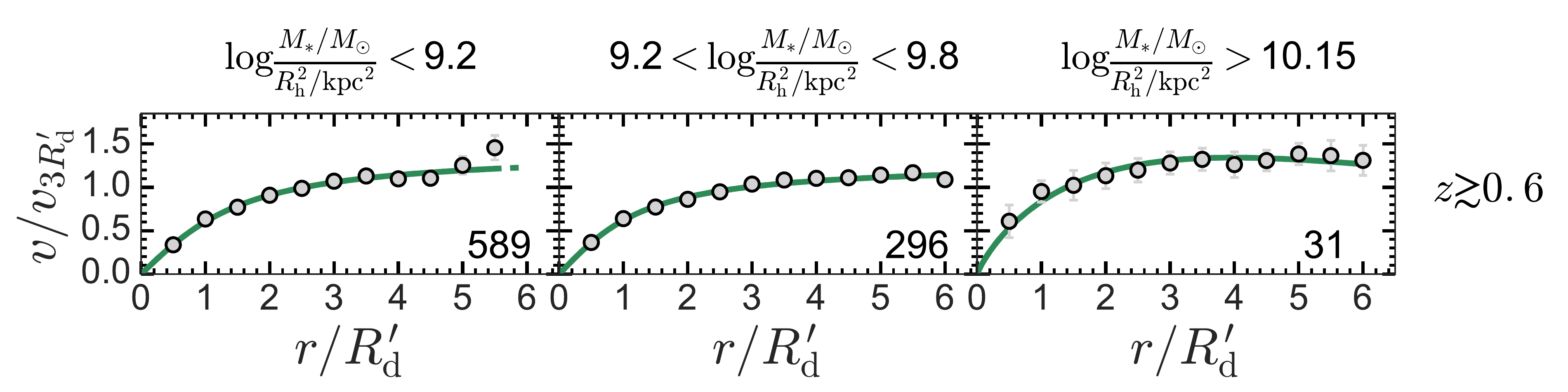}
\end{minipage}
\caption{%
Average wrapped, normalised rotation curves for $z\gtrsim0.6$ galaxies observed with KMOS, separated in to bins of stellar mass surface density. The rotation curves were constructed in the same manner as for Figure~\ref{fig:total_wrapped_curves}. The green solid line in each panel represents the best fit model curve. The line is dashed where the model extrapolates beyond the data. The rotation curves generally become flatter with increasing stellar mass surface density.%
     }%
\label{fig:gridplot_combined}
\end{figure*}

In Figure~\ref{fig:gridplot_combined} we present the average rotation curves extracted from the median stacked position-velocity diagrams for $z\gtrsim0.6$ star-forming galaxies in our sample observed with KMOS. We split the combined sample into three stellar mass surface density bins, constructing an average rotation curve for the galaxies in each bin. These curves are presented in Figure~\ref{fig:gridplot_combined}. With increasing stellar surface mass density, the average galaxy rotation curves become flatter i.e.\ the turnover ($t=v_{6R_{\rm{d}}^{\prime}}/v_{3R_{\rm{d}}^{\prime}}$) reduces. This is in agreement, within uncertainties, with the trend of decreasing $t$ with increasing surface density observed in EAGLE. The corresponding turnovers measured for each curve are presented in Figure~\ref{fig:gridplot_eagle_obs}.

\bsp	
\label{lastpage}
\end{document}